\documentclass[fleqn,10pt]{olplainarticle}

\pdfoutput=1 

\usepackage{todonotes}

\ExplSyntaxOn
\RenewDocumentCommand{\texttt}{m}
 {
  \tl_set:Nn \l_nemgathos_upquotes_tl { #1 }
  \tl_replace_all:Nnn \l_nemgathos_upquotes_tl { '' } { \textquotedbl }
  \tl_replace_all:Nnn \l_nemgathos_upquotes_tl { `` } { \textquotedbl }
  \tl_replace_all:Nnn \l_nemgathos_upquotes_tl { ' } { \textquotesingle }
  \tl_replace_all:Nnn \l_nemgathos_upquotes_tl { ` } { \textquotesingle }
  { \ttfamily \tl_use:N \l_nemgathos_upquotes_tl }
 }
\tl_new:N \l_nemgathos_upquotes_tl
\ExplSyntaxOff

\usepackage{hyperref}
 \hypersetup{
     unicode=false,          
     pdftoolbar=false,        
     pdfmenubar=false,        
     pdffitwindow=false,     
     pdfstartview={FitH},    
     pdfauthor={CosimoPBrogi},     
     colorlinks=true,       
     linkcolor=black,          
     citecolor=black,        
 	urlcolor=blue
 }

\usepackage{stmaryrd}
\usepackage{fdsymbol}
\usepackage[T1]{fontenc}
\DeclareUnicodeCharacter{220F}{{\ensuremath{\prod}}}
\DeclareUnicodeCharacter{03C3}{\ensuremath{\sigma}}
\DeclareUnicodeCharacter{2211}{{\ensuremath{\sum}}}
\DeclareUnicodeCharacter{10627}{{\ensuremath{\llparenthesis}}}
\DeclareUnicodeCharacter{10628}{{\ensuremath{\rrparenthesis}}}
\DeclareUnicodeCharacter{10815}{{\ensuremath{\amalg}}}
\DeclareUnicodeCharacter{9679}{{\ensuremath{\bullet}}}
\DeclareUnicodeCharacter{9726}{{\ensuremath{\blacksquare}}}
\DeclareUnicodeCharacter{9725}{{\ensuremath{\square}}}
\DeclareUnicodeCharacter{10226}{{\ensuremath{\circlearrowleft}}}
\DeclareUnicodeCharacter{10227}{{\ensuremath{\circlearrowright}}}
\DeclareUnicodeCharacter{9645}{{\ensuremath{\boxdot}}}
\DeclareUnicodeCharacter{981}{{\ensuremath{\phi}}}
\DeclareUnicodeCharacter{9565}{{\textSFxxvi}}
\DeclareUnicodeCharacter{8803}{{\ensuremath{\mathop {\vbox {{\rlap =} \kern 3.9pt} {\rlap =} \phantom =} }}}
\DeclareUnicodeCharacter{03BB}{{\ensuremath{\lambda}}}
\DeclareUnicodeCharacter{25CF}{{\ensuremath{\bullet}}}
\DeclareUnicodeCharacter{03B1}{{\ensuremath{\alpha}}}
\DeclareUnicodeCharacter{27E6}{{\ensuremath{\llbracket}}}
\DeclareUnicodeCharacter{27E7}{{\ensuremath{\rrbracket}}}
\DeclareUnicodeCharacter{22C6}{{\ensuremath{\star}}}
\DeclareUnicodeCharacter{21A6}{{\ensuremath{\mapsto}}}
\DeclareUnicodeCharacter{2218}{{\ensuremath{\circ}}}
\DeclareUnicodeCharacter{21B7}{{\ensuremath{\curvearrowright}}}
\DeclareUnicodeCharacter{2243}{{\ensuremath{\simeq}}}
\DeclareUnicodeCharacter{2261}{{\ensuremath{\equiv}}}
\DeclareUnicodeCharacter{019B}{{\ensuremath{\lambda}}}

\usepackage{bussproofs}
\usepackage{fvextra}

\usepackage{etoolbox}
\makeatletter
\patchcmd{\@verbatim}
  {\verbatim@font}
  {\verbatim@font\small}
  {}{}
\makeatother

\usepackage[frozencache=true]{minted}
\newminted[code]{coq}{escapeinside=||,fontsize=\small}
\newcommand{\lncode}{\mintinline[escapeinside=||,mathescape=true]{coq}}
\newcommand{\preparalink}[1]{\providecommand\swhl{}\renewcommand\swhl{\href{#1}}}

\title{Universal algebra in UniMath}

\author[1]{Gianluca Amato}
\author[2]{Matteo Calosci}
\author[2]{Marco Maggesi}
\author[3]{Cosimo Perini Brogi}
\affil[1]{University of Chieti-Pescara, Italy}
\affil[2]{University of Florence, Italy}
\affil[3]{IMT School for Advanced Studies Lucca, Italy}


\begin{abstract}
We present our library for Universal Algebra in the UniMath framework dealing with multi-sorted signatures, their algebras, and the basics for equation systems.

We show how to implement term algebras over a signature without resorting to general inductive constructions (currently not allowed in UniMath) still retaining the computational nature of the definition.

We prove that our single sorted ground term algebras are instances of homotopy W-types. From this perspective, the library enriches UniMath with a computationally well-behaved implementation of a class of W-types.

Moreover, we give neat constructions of the univalent categories of algebras and equational algebras by using the formalism of displayed categories, and show that the term algebra over a signature is the initial object of the category of algebras.  

Finally, we showcase the computational relevance of our work by sketching some basic examples from algebra and propositional logic.
\end{abstract}

\AtBeginShipout{\AtBeginShipoutUpperLeft{%
  \put(\dimexpr\paperwidth-2cm\relax,-1.5cm){\makebox[0pt][r]{\framebox{
  This work has been published on MSCS
  \href{https://doi.org/10.1017/S0960129524000367}{doi:10.1017/S0960129524000367}
  }}}%
}}

\begin{document}

\flushbottom
\maketitle
\thispagestyle{empty}

\section{Introduction}

We present an implementation of the basics of universal algebra in univalent foundations within the formal environment of UniMath by~\cite{UniMath}.

Universal algebra aims to identify common patterns and properties that emerge across different algebraic structures, leading to a deeper understanding of algebraic systems as a whole.\footnote{We refer to~\cite{algebraic} for an extensive introduction to the topic.} It has strong connections with categorical reasoning, and finds applications in several areas of computer science (in database theory and formal methods~\citep{van2012algebraic}), mathematical logic (mainly, model theory~\citep{DBLP:books/daglib/0067423}) and cybersecurity (including cryptographic and communication protocols~\citep{DBLP:journals/tit/DolevY83}).

Since in all those situations it is natural to study algebraic structures modulo isomorphism, univalent mathematics is especially suited for formalising universal algebra.

The choice of working within the UniMath environment has appeared natural since it provides a minimalist implementation of univalent type theory. At the same time, the system comes with an extensive repository of mechanised results covering several fields of mathematics. It then opens a wide range of possibilities for future development of our formalisation.

The code surveyed here introduces the central notions concerning multi-sorted signatures.
Formally defining all the basics has required a certain care. In particular, we had to introduce heterogeneous vectors and generalise types involving signatures by introducing (what we called) ``sorted types''.

Having signatures, we then have given the related formalisation of the category of algebras using the notion of a displayed category by~\cite{lmcs:5252} over the category of {sorted} hSets, whose univalence is proven by adapting the strategy used for the univalence of functor categories. The resulting construction is still a modular one, and the resulting proof term is more concise, for sure, than the one obtained by checking that algebras and homomorphisms satisfy the axioms for standard categories.

We encode terms as lists of operation symbols to be thought of as instructions for a stack-based machine. Terms are those lists of symbols that may be virtually executed without generating type errors or stack underflows.

We show that defining terms this way still yields the expected (homotopy) W-type structure in the single-sorted case.
Moreover, we prove that the term algebra over a signature is the initial object in the corresponding category and that, more generally, an algebra of terms over a signature and a set of variables has the desired universal mapping property.

Our formalisation also includes the notion of equations and algebras modelling an equation system associated with a signature; as for the category of algebras, we use the displayed category formalism to construct the univalent category of equational algebras over a given signature $\sigma$ as the full subcategory of algebras over $\sigma$ satisfying an equation system. 

\paragraph*{Revision history.}
This work is an expanded version of a prior conference paper \citep{hott-uf20} that was presented at the Workshop on Homotopy Type Theory/Univalent Foundations (HoTT/UF) in 2020. An intermediate version appeared in the fourth author's PhD thesis~\citep{DBLP:phd/basesearch/Brogi22}. Modifications in the current version encompass a comprehensive overhaul of the presentation along with the addition of new content, most notably, 
the study of the homotopy W-type structure of our term algebras.

\subsection{Goals and methodology}\label{sec:goals}

What we have mechanised is not a mathematical novelty, but our endeavour has some payoffs.

On the practical side, the code introduces in the UniMath library a minimal set of definitions and results that is open to the community of developers for future achievements and formal investigations on the relation between pre-categorical research in general algebraic structures and its subsequent development in, e.g., Lawvere theories.

On the technical side, a peculiar feature of our code is the original implementation of term algebras over a signature.
We provide a detailed account of the full construction of the term algebra from ground up starting from the inductive type of natural numbers and deriving step-by-step the necessary intermediate structures such as (heterogeneous) vectors and lists.  Incidentally, this plays well with the coding convention adopted in the UniMath library: both \lncode{record} and \lncode{inductive} types are avoided to keep the system sound from a foundational/philosophical viewpoint.

Accordingly, one of our main goals has been to make \emph{all} our constructions about terms \emph{evaluable} -- as far as possible -- by the built-in automation mechanisms of the proof assistant.
More precisely, we represent each term using a sequence of function symbols.
This sequence is thought to be executed by a stack machine: Each symbol of arity $n$ pops $n$ elements from the stack and pushes a new element at the top. A term is denoted by a sequence of function symbols that a stack-like machine can execute without type errors and stack underflow, returning a stack with a single element.

This approach led us to prove a recursion and induction principle on terms that are evaluable as a functional term of the formal system. This performance is somehow mandatory when adhering to a general 
constructive and computational approach such as (small scale) reflection~\citep{DBLP:journals/jfrea/Beeson16, Gonthier_Mahboubi_2010}: with our formalised stack machine, we have written in UniMath an implicit algorithm to compute terms over a signature; using our induction principle, we can run it -- so to speak -- \emph{within} the very formal system of UniMath, and use it to reason about terms safely.

Moreover, our methodology sympathises (in a sense) with the so-called Poincar\'e principle of~\cite{DBLP:journals/jsc/BarendregtC01}: our implementation of terms allows us to rely on the very core engine of UniMath when dealing with these formal objects so that whenever we want to handle them, we can focus on the actual demonstrative contents of the formalisation, leaving to the automation behind the computer proof assistant the trivial computational steps involved in the very proof-term.
 
Generally speaking, standard categorical presentations, though perspicaciously elegant in their abstractness, lack specific suitability for computerised mathematics. 
By contrast, our goal is justified by a specific need for methodological coherence -- we just sketched it a few lines above -- when approaching a work in formalisation. Having proof terms that the computational machinery of UniMath practically evaluates as a correctly typed function fits the philosophy and aims of the mechanisation of mathematics better than just giving a formal counterpart of traditional mathematical notions that the computer cannot handle feasibly.

\subsection{Paper outline}
In what follows, we survey all the notions we introduced in our implementation, structured as an informal presentation of the code; next, we proceed with the discussion of some examples of algebraic structures, to conclude then with some words on future and related work.

In detail, the paper is structured as follows:
\begin{itemize} 
\item In Section~\ref{sec:prelim}, we introduce some extensions of the UniMath library that are needed in the rest of our work such as general constructions about lists and (heterogeneous) vectors;
\item In Section~\ref{sec:alg}, we formalize the very basics of universal algebra:
{multi-sorted signatures}, their {algebras}, and {homomorphisms};
\item In Sections~\ref{sec:term}, \ref{sec:induction}, and \ref{sec:free}, we present the main details of our implementation of {terms}, prove that {term algebras} and {free algebras} do have the required universal property -- stated as the contractibility of the type of out-going homomorphisms -- and discuss the practical and methodological relevance of our {induction principle} on terms;
\item In Section~\ref{sec:w}, we discuss how our notion of ground term algebra has the structure of a W-type;
\item In Section~\ref{sec:eq}, we introduce {systems of equations} and {equational algebras} over a signature;
\item In Section \ref{sec:cat}, we sketch the main lines of our constructions of the {categories of algebras and equational algebras};
\item Finally, Section \ref{sec:ex} is devoted to three applications of our implementation, namely: lists (Section \ref{sec:lists}), monoids (Section \ref{sec:grp}), and Tarski's semantics of propositional boolean formulas (Section \ref{sec:tarski}).
\end{itemize}

\section{Surveying the code}
In this section, we present and comment on the main formalisations within our library.
Our code is part of the official UniMath distribution\footnote{Freely available from \url{http://unimath.org}.}
. The revision discussed in this paper is \href{https://archive.softwareheritage.org/swh:1:rel:1db4fad6810e17a5a9dcaa0563a43b566aaca252;origin=https://github.com/amato-gianluca/UniMath;visit=swh:1:snp:c840c96e2c4f3571fd3082b82697cbaaac480f2c}{archived} on \href{https://www.softwareheritage.org}{Software Heritage}.
To improve readability, in what follows, most proofs and technicalities are omitted, even though they are available in our repository and reachable via the blue hyperlinks in the paper.
 
The implementation discussed in the present article consists mainly of the files in the directories
\begin{itemize}
\item \preparalink{https://archive.softwareheritage.org/swh:1:dir:623e4ba17cef9663e7c7f22db5079dcc08dc9c7d;origin=https://github.com/amato-gianluca/UniMath;visit=swh:1:snp:c840c96e2c4f3571fd3082b82697cbaaac480f2c;anchor=swh:1:rel:1db4fad6810e17a5a9dcaa0563a43b566aaca252;path=/UniMath/Algebra/Universal/}\lncode{|\swhl{UniMath/Algebra/Universal}|} for the basics of universal algebra (together with auxiliary definitions and results), and
\item \preparalink{https://archive.softwareheritage.org/swh:1:dir:3c79d052bbeab5f408c5966b8e948be1f74fe07f;origin=https://github.com/amato-gianluca/UniMath;visit=swh:1:snp:c840c96e2c4f3571fd3082b82697cbaaac480f2c;anchor=swh:1:rel:1db4fad6810e17a5a9dcaa0563a43b566aaca252;path=/UniMath/CategoryTheory/categories/Universal_Algebra/}\lncode{|\swhl{UniMath/CategoryTheory/categories/Universal\_Algebra}|} for the categories of algebras and equational algebras over a signature.
\end{itemize}
However, some improvements to the UniMath library not stricly connected to universal algebras have been introduced in the following directories:
\begin{itemize}
\item \preparalink{https://archive.softwareheritage.org/swh:1:dir:f52438d5efe70c0d4f31b3266f80fea2977c8cd8;origin=https://github.com/amato-gianluca/UniMath;visit=swh:1:snp:c840c96e2c4f3571fd3082b82697cbaaac480f2c;anchor=swh:1:rel:1db4fad6810e17a5a9dcaa0563a43b566aaca252;path=/UniMath/Combinatorics/}\lncode{|\swhl{UniMath/Combinatorics}|} for vectors, lists and sets with decidable equality;
\item \preparalink{https://archive.softwareheritage.org/swh:1:dir:74d494b4e77765e8c7586278a06c2bae020a22c5;origin=https://github.com/amato-gianluca/UniMath;visit=swh:1:snp:c840c96e2c4f3571fd3082b82697cbaaac480f2c;anchor=swh:1:rel:1db4fad6810e17a5a9dcaa0563a43b566aaca252;path=/UniMath/Induction/W/}\lncode{|\swhl{UniMath/Induction/W}|} for the basic definitions of homotopy W-types.
\end{itemize}

\begin{figure}
    \centering
    \includegraphics[trim={0cm 2cm 0cm 1cm},clip, width=1\textwidth]{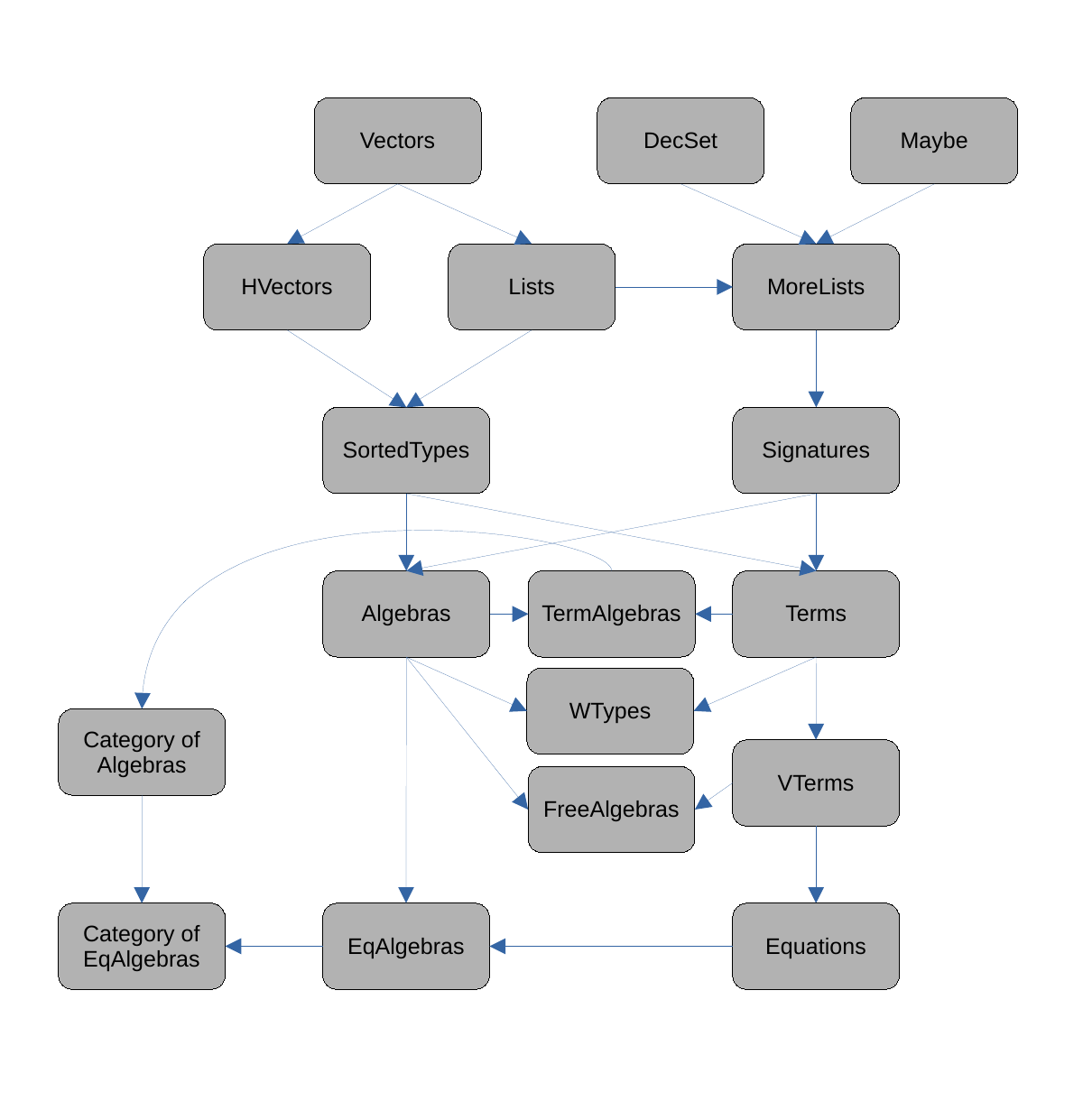}
    \smallskip
    \caption{\label{fig:graph_modules}Intermodule dependencies of the universal algebra formalisation in UniMath.}
\end{figure}

In order to help readers to browse our library, we summarise the dependencies between the files by the diagram in Figure~\ref{fig:graph_modules} -- where an arrow pointing to a node indicate the dependency of the target from the source.

\subsection{Preliminary definitions}\label{sec:prelim}

In order to support the formalization of universal algebra, we have enriched UniMath's standard library with many new concepts and notations. These are introduced in \preparalink{https://archive.softwareheritage.org/swh:1:cnt:07714b3a0abb409b3f355f3e8d5dcd81b4204f48;origin=https://github.com/amato-gianluca/UniMath;visit=swh:1:snp:c840c96e2c4f3571fd3082b82697cbaaac480f2c;anchor=swh:1:rel:1db4fad6810e17a5a9dcaa0563a43b566aaca252;path=/UniMath/Combinatorics/Vectors.v}\lncode{|\swhl{Vectors.v}|}, \preparalink{https://archive.softwareheritage.org/swh:1:cnt:3750de471510aa9d8e821ce739d449c1cc8a78b5;origin=https://github.com/amato-gianluca/UniMath;visit=swh:1:snp:c840c96e2c4f3571fd3082b82697cbaaac480f2c;anchor=swh:1:rel:1db4fad6810e17a5a9dcaa0563a43b566aaca252;path=/UniMath/Combinatorics/Lists.v}\lncode{|\swhl{Lists.v}|}, \preparalink{https://archive.softwareheritage.org/swh:1:cnt:b8b33b1fc690cff9374099fde667a95370980b1f;origin=https://github.com/amato-gianluca/UniMath;visit=swh:1:snp:c840c96e2c4f3571fd3082b82697cbaaac480f2c;anchor=swh:1:rel:1db4fad6810e17a5a9dcaa0563a43b566aaca252;path=/UniMath/Combinatorics/MoreLists.v}\lncode{|\swhl{MoreLists.v}|}, \preparalink{https://archive.softwareheritage.org/swh:1:cnt:d66adf476af0590fc10873027c24c742b1160125;origin=https://github.com/amato-gianluca/UniMath;visit=swh:1:snp:c840c96e2c4f3571fd3082b82697cbaaac480f2c;anchor=swh:1:rel:1db4fad6810e17a5a9dcaa0563a43b566aaca252;path=/UniMath/Algebra/Universal/SortedTypes.v}\lncode{|\swhl{SortedTypes.v}|} and \preparalink{https://archive.softwareheritage.org/swh:1:cnt:4c0091c6bb90c8b3f0c11d8ecfb76967bcd11cda;origin=https://github.com/amato-gianluca/UniMath;visit=swh:1:snp:c840c96e2c4f3571fd3082b82697cbaaac480f2c;anchor=swh:1:rel:1db4fad6810e17a5a9dcaa0563a43b566aaca252;path=/UniMath/Algebra/Universal/HVectors.v}\lncode{|\swhl{HVectors.v}|} files.

\smallskip

The file \lncode{Vectors.v} contains our implementation of the datatype \preparalink{https://archive.softwareheritage.org/swh:1:cnt:07714b3a0abb409b3f355f3e8d5dcd81b4204f48;origin=https://github.com/amato-gianluca/UniMath;visit=swh:1:snp:c840c96e2c4f3571fd3082b82697cbaaac480f2c;anchor=swh:1:rel:1db4fad6810e17a5a9dcaa0563a43b566aaca252;path=/UniMath/Combinatorics/Vectors.v;lines=21-26}\lncode{|\swhl{vec}|} for homogeneous vectors of fixed length.
We changed the implementation of \preparalink{https://archive.softwareheritage.org/swh:1:cnt:3750de471510aa9d8e821ce739d449c1cc8a78b5;origin=https://github.com/amato-gianluca/UniMath;visit=swh:1:snp:c840c96e2c4f3571fd3082b82697cbaaac480f2c;anchor=swh:1:rel:1db4fad6810e17a5a9dcaa0563a43b566aaca252;path=/UniMath/Combinatorics/Lists.v;lines=20-28}\lncode{|\swhl{list}|} in \lncode{Lists.v} in two aspects. First of all, we redefined lists in terms of the new datatype \preparalink{https://archive.softwareheritage.org/swh:1:cnt:07714b3a0abb409b3f355f3e8d5dcd81b4204f48;origin=https://github.com/amato-gianluca/UniMath;visit=swh:1:snp:c840c96e2c4f3571fd3082b82697cbaaac480f2c;anchor=swh:1:rel:1db4fad6810e17a5a9dcaa0563a43b566aaca252;path=/UniMath/Combinatorics/Vectors.v;lines=21-26}\lncode{|\swhl{vec}|} instead of using the ad-hoc type \lncode{iterprod} in the standard version of the file.  Moreover, we changed a couple of theorems from opaque (\lncode{Qed.} conclusion) to transparent (\lncode{Defined.} conclusion). The latter changes are needed to make terms \emph{compute} correctly.
The file \lncode{MoreLists.v} contains notations for lists, such as  \lncode{[v1; ...; vn]} for list literals and \lncode{::} for \emph{cons},  together with additional properties which cannot be found in the standard library.

\smallskip

The type \preparalink{https://archive.softwareheritage.org/swh:1:cnt:4c0091c6bb90c8b3f0c11d8ecfb76967bcd11cda;origin=https://github.com/amato-gianluca/UniMath;visit=swh:1:snp:c840c96e2c4f3571fd3082b82697cbaaac480f2c;anchor=swh:1:rel:1db4fad6810e17a5a9dcaa0563a43b566aaca252;path=/UniMath/Algebra/Universal/HVectors.v;lines=20-24}\lncode{|\swhl{hvec}|} in \lncode{HVectors.v} denotes heterogeneous vectors:\footnote{We need this type to handle operations taking inputs of different sorts. We prefer them to functions since they have better computational properties.} if \lncode{v} is a vector of types \lncode{U1}, \lncode{U2},..., \lncode{Un}, then \lncode{hvec v} is the product type \lncode{U1 × (U2 × ... × (Un × unit))}.
We introduce several basic operations on heterogeneous vectors. Often they have the same syntax as the corresponding operations on plain vectors, and a name which begins with the prefix \lncode{h}. We also introduce notations for heterogeneous vectors, such as \lncode{[(v1; ...; vn)]} for a literal and \lncode{:::} for prefixing. 

\smallskip

Sorted types are types indexed by elements of another type (the index type), so that an element of \preparalink{https://archive.softwareheritage.org/swh:1:cnt:d66adf476af0590fc10873027c24c742b1160125;origin=https://github.com/amato-gianluca/UniMath;visit=swh:1:snp:c840c96e2c4f3571fd3082b82697cbaaac480f2c;anchor=swh:1:rel:1db4fad6810e17a5a9dcaa0563a43b566aaca252;path=/UniMath/Algebra/Universal/SortedTypes.v;lines=20-22}\lncode{|\swhl{sUU}| S} is an \lncode{S}-sorted type, i.e.~an \lncode{S}-indexed family of types.
For functions, \preparalink{https://archive.softwareheritage.org/swh:1:cnt:d66adf476af0590fc10873027c24c742b1160125;origin=https://github.com/amato-gianluca/UniMath;visit=swh:1:snp:c840c96e2c4f3571fd3082b82697cbaaac480f2c;anchor=swh:1:rel:1db4fad6810e17a5a9dcaa0563a43b566aaca252;path=/UniMath/Algebra/Universal/SortedTypes.v;lines=24-29}\lncode{X |\swhl{s→}| Y} denotes the type of \lncode{S}-sorted mapping between \lncode{X} and \lncode{Y}, i.e.~of \lncode{S}-indexed families of functions \lncode{X s → Y s}.

More prominently, for any \lncode{S}-sorted type \lncode{X}, its lifting to \lncode{list S} is denoted by \preparalink{https://archive.softwareheritage.org/swh:1:cnt:d66adf476af0590fc10873027c24c742b1160125;origin=https://github.com/amato-gianluca/UniMath;visit=swh:1:snp:c840c96e2c4f3571fd3082b82697cbaaac480f2c;anchor=swh:1:rel:1db4fad6810e17a5a9dcaa0563a43b566aaca252;path=/UniMath/Algebra/Universal/SortedTypes.v;lines=68-75}\lncode{X|\swhl{*}|}, and is ruled by the identity \lncode{X [s1; s2; ...; sn] = [X s1 ; X s2 ; ... ; X sn]}. Accordingly, if \lncode{f} is an indexed mapping between \lncode{S}-indexed types \lncode{X} and \lncode{Y}, then \preparalink{https://archive.softwareheritage.org/swh:1:cnt:d66adf476af0590fc10873027c24c742b1160125;origin=https://github.com/amato-gianluca/UniMath;visit=swh:1:snp:c840c96e2c4f3571fd3082b82697cbaaac480f2c;anchor=swh:1:rel:1db4fad6810e17a5a9dcaa0563a43b566aaca252;path=/UniMath/Algebra/Universal/SortedTypes.v;lines=77-83}\lncode{f|\swhl{⋆⋆}|} is the lifting of \lncode{f} to a \lncode{list S}-indexed mapping between \lncode{X⋆} and \lncode{Y⋆}. This operation \lncode{⋆⋆} is indeed functorial, and we \preparalink{https://archive.softwareheritage.org/swh:1:cnt:d66adf476af0590fc10873027c24c742b1160125;origin=https://github.com/amato-gianluca/UniMath;visit=swh:1:snp:c840c96e2c4f3571fd3082b82697cbaaac480f2c;anchor=swh:1:rel:1db4fad6810e17a5a9dcaa0563a43b566aaca252;path=/UniMath/Algebra/Universal/SortedTypes.v;lines=89-100}\swhl{prove that} in a form which does not require function extensionality, since resorting to axioms would break computability of terms.

\subsection{Signatures and algebras}\label{sec:alg}

We start by defining a \textbf{multi-sorted signature} to be made of a \emph{decidable set} of sorts along with operations classified by arities and result sorts, as in standard practice.
\preparalink{https://archive.softwareheritage.org/swh:1:cnt:67606886fd74fd78a1dcda9621b9a26f0e0b2801;origin=https://github.com/amato-gianluca/UniMath;visit=swh:1:snp:c840c96e2c4f3571fd3082b82697cbaaac480f2c;anchor=swh:1:rel:1db4fad6810e17a5a9dcaa0563a43b566aaca252;path=/UniMath/Algebra/Universal/Signatures.v;lines=20}
\begin{code}
  Definition |\swhl{signature}| : UU := ∑ (S: decSet) (O: hSet), O → list S × S.
\end{code}
Given \lncode{|$\sigma$|: signature}, we introduce the projections \preparalink{https://archive.softwareheritage.org/swh:1:cnt:67606886fd74fd78a1dcda9621b9a26f0e0b2801;origin=https://github.com/amato-gianluca/UniMath;visit=swh:1:snp:c840c96e2c4f3571fd3082b82697cbaaac480f2c;anchor=swh:1:rel:1db4fad6810e17a5a9dcaa0563a43b566aaca252;path=/UniMath/Algebra/Universal/Signatures.v;lines=22}\lncode{|\swhl{sorts}| |$\sigma$| : decSet} to denote the set of its sorts and
\preparalink{https://archive.softwareheritage.org/swh:1:cnt:67606886fd74fd78a1dcda9621b9a26f0e0b2801;origin=https://github.com/amato-gianluca/UniMath;visit=swh:1:snp:c840c96e2c4f3571fd3082b82697cbaaac480f2c;anchor=swh:1:rel:1db4fad6810e17a5a9dcaa0563a43b566aaca252;path=/UniMath/Algebra/Universal/Signatures.v;lines=24}\lncode{|\swhl{names}| |$\sigma$| : hSet} to denote its set of operations' names. If \lncode{nm : names |$\sigma$|} is also given, then \sloppy 
\preparalink{https://archive.softwareheritage.org/swh:1:cnt:67606886fd74fd78a1dcda9621b9a26f0e0b2801;origin=https://github.com/amato-gianluca/UniMath;visit=swh:1:snp:c840c96e2c4f3571fd3082b82697cbaaac480f2c;anchor=swh:1:rel:1db4fad6810e17a5a9dcaa0563a43b566aaca252;path=/UniMath/Algebra/Universal/Signatures.v;lines=26}\lncode{|\swhl{ar}| |$\sigma$| nm : list S × S} represents its arguments and output sorts. These can also be accessed separately with \preparalink{https://archive.softwareheritage.org/swh:1:cnt:67606886fd74fd78a1dcda9621b9a26f0e0b2801;origin=https://github.com/amato-gianluca/UniMath;visit=swh:1:snp:c840c96e2c4f3571fd3082b82697cbaaac480f2c;anchor=swh:1:rel:1db4fad6810e17a5a9dcaa0563a43b566aaca252;path=/UniMath/Algebra/Universal/Signatures.v;lines=30}\lncode{|\swhl{sort}| nm : sorts |$\sigma$|} and \preparalink{https://archive.softwareheritage.org/swh:1:cnt:67606886fd74fd78a1dcda9621b9a26f0e0b2801;origin=https://github.com/amato-gianluca/UniMath;visit=swh:1:snp:c840c96e2c4f3571fd3082b82697cbaaac480f2c;anchor=swh:1:rel:1db4fad6810e17a5a9dcaa0563a43b566aaca252;path=/UniMath/Algebra/Universal/Signatures.v;lines=28}\lncode{|\swhl{arity}| nm : list (sorts |$\sigma$|)}.

Note that, in a signature, the set of sorts should be a \preparalink{https://archive.softwareheritage.org/swh:1:cnt:43c8f552fc84d67c55bab5204a3b3810aa5882cd;origin=https://github.com/amato-gianluca/UniMath;visit=swh:1:snp:c840c96e2c4f3571fd3082b82697cbaaac480f2c;anchor=swh:1:rel:1db4fad6810e17a5a9dcaa0563a43b566aaca252;path=/UniMath/Combinatorics/DecSet.v;lines=10}\lncode{|\swhl{decSet}|}: this is a type whose equality is decidable, as defined in the file \preparalink{https://archive.softwareheritage.org/swh:1:cnt:43c8f552fc84d67c55bab5204a3b3810aa5882cd;origin=https://github.com/amato-gianluca/UniMath;visit=swh:1:snp:c840c96e2c4f3571fd3082b82697cbaaac480f2c;anchor=swh:1:rel:1db4fad6810e17a5a9dcaa0563a43b566aaca252;path=/UniMath/Combinatorics/DecSet.v}\lncode{|\swhl{DecSet.v}|}. We need this extra property because -- as we previously stated -- we want to \emph{evaluate} terms in the UniMath engine: we can achieve that by pushing sorts into a stack, and we need to check that the very stack contains certain sequences of sorts before applying an operator symbol. Note also that a \lncode{decSet} enjoys the defining property of an \lncode{hSet}. Operators are only required to be in \lncode{hSet}.

A signature may be alternatively specified through the type \preparalink{https://archive.softwareheritage.org/swh:1:cnt:67606886fd74fd78a1dcda9621b9a26f0e0b2801;origin=https://github.com/amato-gianluca/UniMath;visit=swh:1:snp:c840c96e2c4f3571fd3082b82697cbaaac480f2c;anchor=swh:1:rel:1db4fad6810e17a5a9dcaa0563a43b566aaca252;path=/UniMath/Algebra/Universal/Signatures.v;lines=50-56}\lncode{|\swhl{signature\_simple}|}. In a simple signature, the types for sorts and operation symbols are standard finite sets, and the map from operation symbols to domain and range is replaced by a list.
In this way, the definition of a new signature is made simpler.
\begin{code}
  Definition signature_simple : UU := ∑ (ns: nat), list (list (⟦ ns ⟧) × ⟦ ns ⟧).
  
  Definition make_signature_simple {ns: nat} (ar: list (list (⟦ ns ⟧) × ⟦ ns ⟧))
    : signature_simple := ns ,, ar.
    
  Coercion signature_simple_compile (σ: signature_simple) : signature
    := make_signature (⟦ pr1 σ ⟧ ,, isdeceqstn _)
                      (stnset (length (pr2 σ))) (nth (pr2 σ)).
\end{code}
\emph{Single-sorted signatures} are then defined as special cases of \lncode{signature_simple}.
\preparalink{https://archive.softwareheritage.org/swh:1:cnt:67606886fd74fd78a1dcda9621b9a26f0e0b2801;origin=https://github.com/amato-gianluca/UniMath;visit=swh:1:snp:c840c96e2c4f3571fd3082b82697cbaaac480f2c;anchor=swh:1:rel:1db4fad6810e17a5a9dcaa0563a43b566aaca252;path=/UniMath/Algebra/Universal/Signatures.v;lines=58-63}
\begin{code}
  Definition |\swhl{signature\_simple\_single\_sorted}| : UU := list nat.
  
  Definition make_signature_simple_single_sorted (ar: list nat) : 
  	signature_simple_single_sorted := ar.
   
  Coercion signature_simple_single_sorted_compile
      (σ: signature_simple_single_sorted)
    : signature
    := make_signature_single_sorted (stnset (length σ)) (nth σ).
\end{code}
Moving to the file \preparalink{https://archive.softwareheritage.org/swh:1:cnt:3a898cb11b366ae4c72e95545d09cdadcc20534a;origin=https://github.com/amato-gianluca/UniMath;visit=swh:1:snp:c840c96e2c4f3571fd3082b82697cbaaac480f2c;anchor=swh:1:rel:1db4fad6810e17a5a9dcaa0563a43b566aaca252;path=/UniMath/Algebra/Universal/Algebras.v}\lncode{|\swhl{Algebras.v}|}, we define an \textbf{algebra} over a given signature $\sigma$ to be, as usual, support types indexed by sorts together with operations with appropriate sorts:

\preparalink{https://archive.softwareheritage.org/swh:1:cnt:3a898cb11b366ae4c72e95545d09cdadcc20534a;origin=https://github.com/amato-gianluca/UniMath;visit=swh:1:snp:c840c96e2c4f3571fd3082b82697cbaaac480f2c;anchor=swh:1:rel:1db4fad6810e17a5a9dcaa0563a43b566aaca252;path=/UniMath/Algebra/Universal/Algebras.v;lines=13-14}
\begin{code}
  Definition |\swhl{algebra}| (σ: signature): UU
    := ∑ A: sUU (sorts σ), ∏ nm: names σ, A⋆ (arity nm) → A (sort nm).
\end{code}

Given an algebra \lncode{A: algebra σ} we can access its underlying types with \preparalink{https://archive.softwareheritage.org/swh:1:cnt:3a898cb11b366ae4c72e95545d09cdadcc20534a;origin=https://github.com/amato-gianluca/UniMath;visit=swh:1:snp:c840c96e2c4f3571fd3082b82697cbaaac480f2c;anchor=swh:1:rel:1db4fad6810e17a5a9dcaa0563a43b566aaca252;path=/UniMath/Algebra/Universal/Algebras.v;lines=16-18}\lncode{|\swhl{support}| A : sUU (sorts σ)} and its operations with \preparalink{https://archive.softwareheritage.org/swh:1:cnt:3a898cb11b366ae4c72e95545d09cdadcc20534a;origin=https://github.com/amato-gianluca/UniMath;visit=swh:1:snp:c840c96e2c4f3571fd3082b82697cbaaac480f2c;anchor=swh:1:rel:1db4fad6810e17a5a9dcaa0563a43b566aaca252;path=/UniMath/Algebra/Universal/Algebras.v;lines=20}\lncode{|\swhl{ops}| A}. If the name \lncode{nm} of such an operation is given then we can access the domain and range of the corresponding operation as interpreted in \lncode{A} with \preparalink{https://archive.softwareheritage.org/swh:1:cnt:3a898cb11b366ae4c72e95545d09cdadcc20534a;origin=https://github.com/amato-gianluca/UniMath;visit=swh:1:snp:c840c96e2c4f3571fd3082b82697cbaaac480f2c;anchor=swh:1:rel:1db4fad6810e17a5a9dcaa0563a43b566aaca252;path=/UniMath/Algebra/Universal/Algebras.v;lines=25}\lncode{|\swhl{dom}| A nm : UU} and \preparalink{https://archive.softwareheritage.org/swh:1:cnt:3a898cb11b366ae4c72e95545d09cdadcc20534a;origin=https://github.com/amato-gianluca/UniMath;visit=swh:1:snp:c840c96e2c4f3571fd3082b82697cbaaac480f2c;anchor=swh:1:rel:1db4fad6810e17a5a9dcaa0563a43b566aaca252;path=/UniMath/Algebra/Universal/Algebras.v;lines=27}\lncode{|\swhl{rng}| A nm : UU} respectively.

We declare the projection \lncode{support} as a type coercion. Moreover, as for signatures, we simplify the building term for algebras when starting from a simple signature:
\preparalink{https://archive.softwareheritage.org/swh:1:cnt:3a898cb11b366ae4c72e95545d09cdadcc20534a;origin=https://github.com/amato-gianluca/UniMath;visit=swh:1:snp:c840c96e2c4f3571fd3082b82697cbaaac480f2c;anchor=swh:1:rel:1db4fad6810e17a5a9dcaa0563a43b566aaca252;path=/UniMath/Algebra/Universal/Algebras.v;lines=61-87}
\begin{code}
  Definition |\swhl{make\_algebra\_simple}|
      (σ: signature_simple) (A: vec UU (pr1 σ))
      (ops: (λ a, (el A)⋆ (dirprod_pr1 a) → el A (dirprod_pr2 a))⋆ (pr2 σ))
    : algebra σ.
\end{code}
A similar proof-term (\preparalink{https://archive.softwareheritage.org/swh:1:cnt:3a898cb11b366ae4c72e95545d09cdadcc20534a;origin=https://github.com/amato-gianluca/UniMath;visit=swh:1:snp:c840c96e2c4f3571fd3082b82697cbaaac480f2c;anchor=swh:1:rel:1db4fad6810e17a5a9dcaa0563a43b566aaca252;path=/UniMath/Algebra/Universal/Algebras.v;lines=38-59}\lncode{|\swhl{make\_algebra\_simple\_single\_sorted}|}) is given for single-sorted signatures.

All of these notions allow us to define \textbf{algebra homomorphisms}:
\preparalink{https://archive.softwareheritage.org/swh:1:cnt:3a898cb11b366ae4c72e95545d09cdadcc20534a;origin=https://github.com/amato-gianluca/UniMath;visit=swh:1:snp:c840c96e2c4f3571fd3082b82697cbaaac480f2c;anchor=swh:1:rel:1db4fad6810e17a5a9dcaa0563a43b566aaca252;path=/UniMath/Algebra/Universal/Algebras.v;lines=91-92}
\begin{code}
  Definition |\swhl{ishom}| {σ: signature} {A1 A2: algebra σ} (h: A1 s→ A2) : UU
    := ∏ (nm: names σ) (x: dom A1 nm), h _ (ops A1 nm x) = ops A2 nm (h⋆⋆ _ x).
\end{code}
\preparalink{https://archive.softwareheritage.org/swh:1:cnt:3a898cb11b366ae4c72e95545d09cdadcc20534a;origin=https://github.com/amato-gianluca/UniMath;visit=swh:1:snp:c840c96e2c4f3571fd3082b82697cbaaac480f2c;anchor=swh:1:rel:1db4fad6810e17a5a9dcaa0563a43b566aaca252;path=/UniMath/Algebra/Universal/Algebras.v;lines=94}
\begin{code}
  Definition |\swhl{hom}| {σ: signature} (A1 A2: algebra σ): UU := ∑ (h: A1 s→ A2), ishom h.
\end{code}
The notation  \preparalink{https://archive.softwareheritage.org/swh:1:cnt:3a898cb11b366ae4c72e95545d09cdadcc20534a;origin=https://github.com/amato-gianluca/UniMath;visit=swh:1:snp:c840c96e2c4f3571fd3082b82697cbaaac480f2c;anchor=swh:1:rel:1db4fad6810e17a5a9dcaa0563a43b566aaca252;path=/UniMath/Algebra/Universal/Algebras.v;lines=98}\lncode{A1 |\swhl{↷}| A2} is also introduced as an alternative form for ``\lncode{hom A1 A2}''. A special case is when the support of the target algebra \lncode{A2} is comprised of sets, i.e. when we have an inhabitant of
\preparalink{https://archive.softwareheritage.org/swh:1:cnt:3a898cb11b366ae4c72e95545d09cdadcc20534a;origin=https://github.com/amato-gianluca/UniMath;visit=swh:1:snp:c840c96e2c4f3571fd3082b82697cbaaac480f2c;anchor=swh:1:rel:1db4fad6810e17a5a9dcaa0563a43b566aaca252;path=/UniMath/Algebra/Universal/Algebras.v;lines=29-30}
\begin{code}
  Definition |\swhl{has\_supportsets}| {σ: signature} (A: algebra σ): UU
    := ∏ s: sorts σ, isaset (support A s).
\end{code}
In this case, \lncode|ishom| is a property and \lncode|A1 ↷ A2| is a set:
\preparalink{https://archive.softwareheritage.org/swh:1:cnt:3a898cb11b366ae4c72e95545d09cdadcc20534a;origin=https://github.com/amato-gianluca/UniMath;visit=swh:1:snp:c840c96e2c4f3571fd3082b82697cbaaac480f2c;anchor=swh:1:rel:1db4fad6810e17a5a9dcaa0563a43b566aaca252;path=/UniMath/Algebra/Universal/Algebras.v;lines=114-123}
\begin{code}
  Theorem |\swhl{isapropishom}| {σ: signature} {A1 A2: algebra σ} (f: sfun A1 A2)
    (setprop: has_supportsets A2) : isaprop (ishom f).
\end{code}
\preparalink{https://archive.softwareheritage.org/swh:1:cnt:3a898cb11b366ae4c72e95545d09cdadcc20534a;origin=https://github.com/amato-gianluca/UniMath;visit=swh:1:snp:c840c96e2c4f3571fd3082b82697cbaaac480f2c;anchor=swh:1:rel:1db4fad6810e17a5a9dcaa0563a43b566aaca252;path=/UniMath/Algebra/Universal/Algebras.v;lines=125-139}
\begin{code}
  Theorem |\swhl{isasethom}| {σ: signature} (A1 A2: algebra σ)
    (setprop: has_supportsets A2) :  isaset (A1 ↷ A2).
\end{code}
Next, we prove -- by lemmas \preparalink{https://archive.softwareheritage.org/swh:1:cnt:3a898cb11b366ae4c72e95545d09cdadcc20534a;origin=https://github.com/amato-gianluca/UniMath;visit=swh:1:snp:c840c96e2c4f3571fd3082b82697cbaaac480f2c;anchor=swh:1:rel:1db4fad6810e17a5a9dcaa0563a43b566aaca252;path=/UniMath/Algebra/Universal/Algebras.v;lines=143-151}\lncode{|\swhl{ishomid}|} and \preparalink{https://archive.softwareheritage.org/swh:1:cnt:3a898cb11b366ae4c72e95545d09cdadcc20534a;origin=https://github.com/amato-gianluca/UniMath;visit=swh:1:snp:c840c96e2c4f3571fd3082b82697cbaaac480f2c;anchor=swh:1:rel:1db4fad6810e17a5a9dcaa0563a43b566aaca252;path=/UniMath/Algebra/Universal/Algebras.v;lines=153-167}\lncode{|\swhl{ishomcomp}|} -- that the identity function determines an identity homomorphism, and that the property \lncode{ishom} is closed under composition.

\subsection{Terms and free algebras}\label{sec:term}

The file \preparalink{https://archive.softwareheritage.org/swh:1:cnt:3a898cb11b366ae4c72e95545d09cdadcc20534a;origin=https://github.com/amato-gianluca/UniMath;visit=swh:1:snp:c840c96e2c4f3571fd3082b82697cbaaac480f2c;anchor=swh:1:rel:1db4fad6810e17a5a9dcaa0563a43b566aaca252;path=/UniMath/Algebra/Universal/Algebras.v}\lncode{|\swhl{Algebras.v}|} is closed by the construction of the unit algebra and a proof of its finality among those defined over its signature:
\preparalink{https://archive.softwareheritage.org/swh:1:cnt:3a898cb11b366ae4c72e95545d09cdadcc20534a;origin=https://github.com/amato-gianluca/UniMath;visit=swh:1:snp:c840c96e2c4f3571fd3082b82697cbaaac480f2c;anchor=swh:1:rel:1db4fad6810e17a5a9dcaa0563a43b566aaca252;path=/UniMath/Algebra/Universal/Algebras.v;lines=171-172}
\begin{code}
  Definition |\swhl{unitalgebra}| (σ: signature): algebra σ
    := make_algebra (sunit (sorts σ)) tosunit.
\end{code}
\preparalink{https://archive.softwareheritage.org/swh:1:cnt:3a898cb11b366ae4c72e95545d09cdadcc20534a;origin=https://github.com/amato-gianluca/UniMath;visit=swh:1:snp:c840c96e2c4f3571fd3082b82697cbaaac480f2c;anchor=swh:1:rel:1db4fad6810e17a5a9dcaa0563a43b566aaca252;path=/UniMath/Algebra/Universal/Algebras.v;lines=184-195}
\begin{code}
  Theorem |\swhl{iscontrhomstounit}| {σ: signature} (A: algebra σ)
    : iscontr (hom A (unitalgebra σ)).
\end{code}
However, we are mostly interested in the \emph{initial} object of the category of algebras, namely the algebra of terms over a given signature. In standard 
textbooks, the set of terms over a signature $\sigma$ and a (disjoint) set $V$ of variables is defined as the least set including $V$ and closed under application of symbols of $\sigma$.

Without recurring to general inductive types, in \preparalink{https://archive.softwareheritage.org/swh:1:cnt:78b995258734161c8d59e5e8ec99ca7293d55de4;origin=https://github.com/amato-gianluca/UniMath;visit=swh:1:snp:c840c96e2c4f3571fd3082b82697cbaaac480f2c;anchor=swh:1:rel:1db4fad6810e17a5a9dcaa0563a43b566aaca252;path=/UniMath/Algebra/Universal/Terms.v}\lncode{|\swhl{Terms.v}|} we implement this notion using an alternative device, based on reverse Polish notation and value stacks. 

In our formalisation we start with the special case where the set of variables $V$ is empty.
The rough and general idea can be sketched as follows:
\begin{enumerate}
\item A sequence of function symbols is thought of as a series of commands to be executed by a \emph{stack machine} whose stack is made of sorts, and which we define by means of a maybe monad we construct from raw in \preparalink{https://archive.softwareheritage.org/swh:1:cnt:51042b733460e2f2a06670aa4f4be4d085d7b4e6;origin=https://github.com/amato-gianluca/UniMath;visit=swh:1:snp:c840c96e2c4f3571fd3082b82697cbaaac480f2c;anchor=swh:1:rel:1db4fad6810e17a5a9dcaa0563a43b566aaca252;path=/UniMath/Combinatorics/Maybe.v}\lncode{|\swhl{Maybe.v}|}.
\preparalink{https://archive.softwareheritage.org/swh:1:cnt:78b995258734161c8d59e5e8ec99ca7293d55de4;origin=https://github.com/amato-gianluca/UniMath;visit=swh:1:snp:c840c96e2c4f3571fd3082b82697cbaaac480f2c;anchor=swh:1:rel:1db4fad6810e17a5a9dcaa0563a43b566aaca252;path=/UniMath/Algebra/Universal/Terms.v;lines=36-40}
\begin{code}
  Local Definition |\swhl{oplist}| (σ: signature):= list (names σ).
\end{code}
\preparalink{https://archive.softwareheritage.org/swh:1:cnt:78b995258734161c8d59e5e8ec99ca7293d55de4;origin=https://github.com/amato-gianluca/UniMath;visit=swh:1:snp:c840c96e2c4f3571fd3082b82697cbaaac480f2c;anchor=swh:1:rel:1db4fad6810e17a5a9dcaa0563a43b566aaca252;path=/UniMath/Algebra/Universal/Terms.v;lines=48}
\begin{code}
  Local Definition |\swhl{stack}| (σ: signature): UU := maybe (list (sorts σ)).
\end{code}

\item When an operation symbol is executed, its arity is popped out from the stack and replaced by its range. When a stack underflow occurs, or when the sorts present in the stack are not the ones expected by the operator, the stack goes into an error condition which is propagated by successive operations. We implement this process by means of two functions:
\begin{code}
  Local Definition opexec (nm: names σ): stack σ → stack σ
    := flatmap (λ ss, just (sort nm :: ss)) ∘
       flatmap (λ ss, prefix_remove (arity nm) ss).
\end{code}
\begin{code}
  Local Definition oplistexec (l: oplist σ): stack σ := foldr opexec (just []) l.
\end{code}

The former is the stack transformation corresponding to the execution of the operation symbol \lncode{nm}. The latter returns the stack corresponding to the
execution of the entire \lncode{oplist} argument starting from the empty stack. The list is executed from the last to the first operation symbol.

Several additional lemmas are required in order to make us able to handle stacks -- by concatenating, splitting, etc.~-- without incurring failures breaking down the whole process.\footnote{In particular, since we need to decide when a stack is correctly executed and when an underflow occurs, we see the reasons for choosing sorts to constitute a decidable set.}
\item Finally, we define a term to be just a list of operation symbols that, after being executed by \lncode{oplistexec}, returns a list of length one with appropriate sort:\footnote{From a purely HoTT-perspective, we can easily see also that the type of stacks over $\sigma$ is an hSet, so that the property of being a term is not proof-relevant \preparalink{https://archive.softwareheritage.org/swh:1:cnt:78b995258734161c8d59e5e8ec99ca7293d55de4;origin=https://github.com/amato-gianluca/UniMath;visit=swh:1:snp:c840c96e2c4f3571fd3082b82697cbaaac480f2c;anchor=swh:1:rel:1db4fad6810e17a5a9dcaa0563a43b566aaca252;path=/UniMath/Algebra/Universal/Terms.v;lines=78-81}(\swhl{\texttt{isapropisaterm}}).}
\preparalink{https://archive.softwareheritage.org/swh:1:cnt:78b995258734161c8d59e5e8ec99ca7293d55de4;origin=https://github.com/amato-gianluca/UniMath;visit=swh:1:snp:c840c96e2c4f3571fd3082b82697cbaaac480f2c;anchor=swh:1:rel:1db4fad6810e17a5a9dcaa0563a43b566aaca252;path=/UniMath/Algebra/Universal/Terms.v;lines=76}
\begin{code}
  Local Definition |\swhl{isaterm}| (s: sorts σ) (l: oplist σ): UU
    := oplistexec l = just ([s]).
\end{code}
\preparalink{https://archive.softwareheritage.org/swh:1:cnt:78b995258734161c8d59e5e8ec99ca7293d55de4;origin=https://github.com/amato-gianluca/UniMath;visit=swh:1:snp:c840c96e2c4f3571fd3082b82697cbaaac480f2c;anchor=swh:1:rel:1db4fad6810e17a5a9dcaa0563a43b566aaca252;path=/UniMath/Algebra/Universal/Terms.v;lines=440-443}
\begin{code}
  Local Definition |\swhl{term}| (σ: signature) (s: sorts σ): UU
    := ∑ t: oplist σ, isaterm s t.
\end{code}
\end{enumerate}

\begin{code}
  Local Definition build_term (nm: names σ) (v: (term σ)⋆ (arity nm)):
    term σ (sort nm).
\end{code}
The implementation of \lncode{build_term} is quite straightforward. It concatenates \lncode{nm} and the oplists underlying the terms in \lncode{v}, and builds a proof that the resulting oplist is a term from the proofs that the elements of \lncode{v} are terms. The \lncode{princop} and \lncode{subterms} accessors are projections of a more complex operation called \preparalink{https://archive.softwareheritage.org/swh:1:cnt:78b995258734161c8d59e5e8ec99ca7293d55de4;origin=https://github.com/amato-gianluca/UniMath;visit=swh:1:snp:c840c96e2c4f3571fd3082b82697cbaaac480f2c;anchor=swh:1:rel:1db4fad6810e17a5a9dcaa0563a43b566aaca252;path=/UniMath/Algebra/Universal/Terms.v;lines=612-646}\lncode{|\swhl{term\_decompose}|} which breaks a term in principal operation symbols \lncode{nm} and subterms \lncode{v}, and, at the same time, provides the proof-terms that characterize their behaviour.

\subsection{Induction on terms}
\label{sec:induction}

At this point, we proceed in proving induction over terms. The inductive hypothesis, being quite complex, is stated in the \preparalink{https://archive.softwareheritage.org/swh:1:cnt:78b995258734161c8d59e5e8ec99ca7293d55de4;origin=https://github.com/amato-gianluca/UniMath;visit=swh:1:snp:c840c96e2c4f3571fd3082b82697cbaaac480f2c;anchor=swh:1:rel:1db4fad6810e17a5a9dcaa0563a43b566aaca252;path=/UniMath/Algebra/Universal/Terms.v;lines=740-744}\lncode{|\swhl{term\_ind\_HP}|} type.

\begin{code}
  Definition term_ind_HP (P: ∏ (s: sorts σ), term σ s → UU) : UU
    := ∏ (nm: names σ) (v: (term σ)⋆ (arity nm)) (IH: hvec (h1map_vec P v)),
       P (sort nm) (build_term nm v).
\end{code}
Given a family \lncode{P} of types, indexed by a sort \lncode{s} and a term over \lncode{s}, the inductive hypothesis is a function that, given an operation symbol \lncode{nm}, a sequence of terms \lncode{v}, and a sequence of proofs of \lncode{P} for all terms in \lncode{v}, is able to build a proof of \lncode{P} for the term \lncode{build_term nm v}, i.e.~$nm(v_1, \ldots,v_n)$. The identifier \preparalink{https://archive.softwareheritage.org/swh:1:cnt:4c0091c6bb90c8b3f0c11d8ecfb76967bcd11cda;origin=https://github.com/amato-gianluca/UniMath;visit=swh:1:snp:c840c96e2c4f3571fd3082b82697cbaaac480f2c;anchor=swh:1:rel:1db4fad6810e17a5a9dcaa0563a43b566aaca252;path=/UniMath/Algebra/Universal/HVectors.v;lines=367-369}\lncode{|\swhl{h1map\_vec}|} simply denotes a variant of \preparalink{https://archive.softwareheritage.org/swh:1:cnt:07714b3a0abb409b3f355f3e8d5dcd81b4204f48;origin=https://github.com/amato-gianluca/UniMath;visit=swh:1:snp:c840c96e2c4f3571fd3082b82697cbaaac480f2c;anchor=swh:1:rel:1db4fad6810e17a5a9dcaa0563a43b566aaca252;path=/UniMath/Combinatorics/Vectors.v;lines=216-220}\lncode{|\swhl{vec\_map}|} for heterogeneous vectors.  Given this auxiliary definition, the \textbf{induction principle for terms} may be easily stated as follows:
\preparalink{https://archive.softwareheritage.org/swh:1:cnt:78b995258734161c8d59e5e8ec99ca7293d55de4;origin=https://github.com/amato-gianluca/UniMath;visit=swh:1:snp:c840c96e2c4f3571fd3082b82697cbaaac480f2c;anchor=swh:1:rel:1db4fad6810e17a5a9dcaa0563a43b566aaca252;path=/UniMath/Algebra/Universal/Terms.v;lines=782-786}
\begin{code}
  Theorem |\swhl{term\_ind}| (P: ∏ (s: sorts σ), term σ s → UU) (R: term_ind_HP P) 
                   {s: sorts σ} (t: term σ s)
    : P s t.
\end{code}
The proof proceeds by induction on the length of the oplist underlying \lncode{t}, using the \preparalink{https://archive.softwareheritage.org/swh:1:cnt:78b995258734161c8d59e5e8ec99ca7293d55de4;origin=https://github.com/amato-gianluca/UniMath;visit=swh:1:snp:c840c96e2c4f3571fd3082b82697cbaaac480f2c;anchor=swh:1:rel:1db4fad6810e17a5a9dcaa0563a43b566aaca252;path=/UniMath/Algebra/Universal/Terms.v;lines=746-769}\lncode{|\swhl{term\_ind\_onlength}|} auxiliary function.

Simple examples of use of the induction principle on terms are the \preparalink{https://archive.softwareheritage.org/swh:1:cnt:78b995258734161c8d59e5e8ec99ca7293d55de4;origin=https://github.com/amato-gianluca/UniMath;visit=swh:1:snp:c840c96e2c4f3571fd3082b82697cbaaac480f2c;anchor=swh:1:rel:1db4fad6810e17a5a9dcaa0563a43b566aaca252;path=/UniMath/Algebra/Universal/Terms.v;lines=871-874}\lncode{|\swhl{depth}|} and \preparalink{https://archive.softwareheritage.org/swh:1:cnt:78b995258734161c8d59e5e8ec99ca7293d55de4;origin=https://github.com/amato-gianluca/UniMath;visit=swh:1:snp:c840c96e2c4f3571fd3082b82697cbaaac480f2c;anchor=swh:1:rel:1db4fad6810e17a5a9dcaa0563a43b566aaca252;path=/UniMath/Algebra/Universal/Terms.v;lines=876-878}\lncode{|\swhl{fromterm}|} functions.
The former computes the depth of a term, and the latter is essentially the evaluation map for ground terms in an algebra.
The \lncode{h2lower} proof term which appears in the definition of \lncode{fromterm} is just a technicality needed to convert between types which are provably equal but not convertible. This might be replaced by a transport, if we were not interested in computability. The same can be said for the proof term \lncode{h1lift}, later in the definition of \lncode{term_ind_step}.
\begin{code}
  Local Definition fromterm {A: sUU (sorts σ)} 
                            (op : ∏ (nm : names σ), A⋆ (arity nm) → A (sort nm))
                            {s: sorts σ}
    : term σ s → A s
    := term_ind (λ s _, A s) (λ nm v rec, op nm (h2lower rec)).
\end{code}

In order to reason effectively on inductive definitions, we need an induction unfolding property. For natural numbers, it is
\begin{code}
  nat_rect P a IH (S n) = IH n (nat_rect P a IH n)
\end{code}
which means that the result of applying the recursive definition to \lncode{S n} may be obtained by applying the recursive definition to \lncode{n} and then the inductive hypothesis. While this induction unfolding properties are provable just by \lncode{reflexivity} for many inductive types, this does not hold for terms, and a quite complex proof is needed:
\preparalink{https://archive.softwareheritage.org/swh:1:cnt:78b995258734161c8d59e5e8ec99ca7293d55de4;origin=https://github.com/amato-gianluca/UniMath;visit=swh:1:snp:c840c96e2c4f3571fd3082b82697cbaaac480f2c;anchor=swh:1:rel:1db4fad6810e17a5a9dcaa0563a43b566aaca252;path=/UniMath/Algebra/Universal/Terms.v;lines=830-862}
\begin{code}
  Lemma |\swhl{term\_ind\_step}| (P: ∏ (s: sorts σ), term σ s → UU) (R: term_ind_HP P) 
                      (nm: names σ) (v: (term σ)⋆ (arity nm))
    : term_ind P R (build_term nm v) 
      = R nm v (h2map (λ s t q, term_ind P R t) (h1lift v)).
\end{code}

Notice that many of the definitions which appear in \lncode{Terms.v} are declared as \lncode{Local}. This is so because they are considered internal implementation details and should not be used unless explicitly needed. In particular, this holds for a set of identifiers that will be redefined in \preparalink{https://archive.softwareheritage.org/swh:1:cnt:485d80e1be15fce12cfba0244b95752fd1bf4bb3;origin=https://github.com/amato-gianluca/UniMath;visit=swh:1:snp:c840c96e2c4f3571fd3082b82697cbaaac480f2c;anchor=swh:1:rel:1db4fad6810e17a5a9dcaa0563a43b566aaca252;path=/UniMath/Algebra/Universal/VTerms.v}\lncode{|\swhl{VTerms.v}|} to work on terms with variables. Since sometimes it may be convenient to have specialized functions that only work with ground terms, they are exported through a series of notations, such as:
\preparalink{https://archive.softwareheritage.org/swh:1:cnt:78b995258734161c8d59e5e8ec99ca7293d55de4;origin=https://github.com/amato-gianluca/UniMath;visit=swh:1:snp:c840c96e2c4f3571fd3082b82697cbaaac480f2c;anchor=swh:1:rel:1db4fad6810e17a5a9dcaa0563a43b566aaca252;path=/UniMath/Algebra/Universal/Terms.v;lines=922}
\begin{code}
  Notation |\swhl{gterm}| := term.
\end{code}
\preparalink{https://archive.softwareheritage.org/swh:1:cnt:78b995258734161c8d59e5e8ec99ca7293d55de4;origin=https://github.com/amato-gianluca/UniMath;visit=swh:1:snp:c840c96e2c4f3571fd3082b82697cbaaac480f2c;anchor=swh:1:rel:1db4fad6810e17a5a9dcaa0563a43b566aaca252;path=/UniMath/Algebra/Universal/Terms.v;lines=930}
\begin{code}
  Notation |\swhl{build\_gterm}| := build_term.
\end{code}

\subsection{Terms with variables and free algebras}\label{sec:free}

Considering \textbf{terms with variables} is what we do in file \preparalink{https://archive.softwareheritage.org/swh:1:cnt:485d80e1be15fce12cfba0244b95752fd1bf4bb3;origin=https://github.com/amato-gianluca/UniMath;visit=swh:1:snp:c840c96e2c4f3571fd3082b82697cbaaac480f2c;anchor=swh:1:rel:1db4fad6810e17a5a9dcaa0563a43b566aaca252;path=/UniMath/Algebra/Universal/VTerms.v}\lncode{|\swhl{VTerms.v}|}. The idea is that a term with variables in $V$ over a signature \lncode{σ} is a ground term in a new signature where constant symbols are enlarged with the variables in $V$.
Variables and corresponding sorts are declared in a \preparalink{https://archive.softwareheritage.org/swh:1:cnt:485d80e1be15fce12cfba0244b95752fd1bf4bb3;origin=https://github.com/amato-gianluca/UniMath;visit=swh:1:snp:c840c96e2c4f3571fd3082b82697cbaaac480f2c;anchor=swh:1:rel:1db4fad6810e17a5a9dcaa0563a43b566aaca252;path=/UniMath/Algebra/Universal/VTerms.v;lines=20-23}\lncode{|\swhl{varspec}|} (\emph{variable specification}), while \preparalink{https://archive.softwareheritage.org/swh:1:cnt:485d80e1be15fce12cfba0244b95752fd1bf4bb3;origin=https://github.com/amato-gianluca/UniMath;visit=swh:1:snp:c840c96e2c4f3571fd3082b82697cbaaac480f2c;anchor=swh:1:rel:1db4fad6810e17a5a9dcaa0563a43b566aaca252;path=/UniMath/Algebra/Universal/VTerms.v;lines=29-30}\lncode{|\swhl{vsignature}|} builds the new signature.
\begin{code}
  Definition varspec (σ: signature) := ∑ V: hSet, V → sorts σ.
\end{code}
\begin{code}
  Definition vsignature (σ : signature) (V: varspec σ): signature
    := make_signature (sorts σ) (setcoprod (names σ) V) 
                                (sumofmaps (ar σ) (λ v, nil ,, varsort v)).
\end{code}
The proof-terms \preparalink{https://archive.softwareheritage.org/swh:1:cnt:485d80e1be15fce12cfba0244b95752fd1bf4bb3;origin=https://github.com/amato-gianluca/UniMath;visit=swh:1:snp:c840c96e2c4f3571fd3082b82697cbaaac480f2c;anchor=swh:1:rel:1db4fad6810e17a5a9dcaa0563a43b566aaca252;path=/UniMath/Algebra/Universal/VTerms.v;lines=34}\lncode{|\swhl{namelift}|} and \preparalink{https://archive.softwareheritage.org/swh:1:cnt:485d80e1be15fce12cfba0244b95752fd1bf4bb3;origin=https://github.com/amato-gianluca/UniMath;visit=swh:1:snp:c840c96e2c4f3571fd3082b82697cbaaac480f2c;anchor=swh:1:rel:1db4fad6810e17a5a9dcaa0563a43b566aaca252;path=/UniMath/Algebra/Universal/VTerms.v;lines=36}\lncode{|\swhl{varname}|} are the injections of, respectively, operation sysmbols and variables in the extended signature.

Then, a list of definitions comes: they essentially introduce terms with variables by resorting to ground terms.
\preparalink{https://archive.softwareheritage.org/swh:1:cnt:485d80e1be15fce12cfba0244b95752fd1bf4bb3;origin=https://github.com/amato-gianluca/UniMath;visit=swh:1:snp:c840c96e2c4f3571fd3082b82697cbaaac480f2c;anchor=swh:1:rel:1db4fad6810e17a5a9dcaa0563a43b566aaca252;path=/UniMath/Algebra/Universal/VTerms.v;lines=38}
\begin{code}
  Definition |\swhl{term}| (σ: signature) (V: varspec σ)
    : sUU (sorts σ) := gterm (vsignature σ V).
\end{code}
\preparalink{https://archive.softwareheritage.org/swh:1:cnt:485d80e1be15fce12cfba0244b95752fd1bf4bb3;origin=https://github.com/amato-gianluca/UniMath;visit=swh:1:snp:c840c96e2c4f3571fd3082b82697cbaaac480f2c;anchor=swh:1:rel:1db4fad6810e17a5a9dcaa0563a43b566aaca252;path=/UniMath/Algebra/Universal/VTerms.v;lines=42-43}
\begin{code}
  Definition |\swhl{build\_term}| {V: varspec σ} (nm: names σ) (v: (term σ V)⋆ (arity nm))
    : term σ V (sort nm) := build_gterm (namelift V nm) v.
\end{code}
\preparalink{https://archive.softwareheritage.org/swh:1:cnt:485d80e1be15fce12cfba0244b95752fd1bf4bb3;origin=https://github.com/amato-gianluca/UniMath;visit=swh:1:snp:c840c96e2c4f3571fd3082b82697cbaaac480f2c;anchor=swh:1:rel:1db4fad6810e17a5a9dcaa0563a43b566aaca252;path=/UniMath/Algebra/Universal/VTerms.v;lines=45}
\begin{code}
  Definition |\swhl{varterm}| {V: varspec σ} (v: V)
    : term σ V (varsort v) := build_gterm (varname v) [()].
\end{code}

Finally, in \preparalink{https://archive.softwareheritage.org/swh:1:cnt:267b5bdd29a0bef5905d5289e52708282d8e8323;origin=https://github.com/amato-gianluca/UniMath;visit=swh:1:snp:c840c96e2c4f3571fd3082b82697cbaaac480f2c;anchor=swh:1:rel:1db4fad6810e17a5a9dcaa0563a43b566aaca252;path=/UniMath/Algebra/Universal/FreeAlgebras.v}\lncode{|\swhl{FreeAlgebras.v}|} we pack terms and the \lncode{build_term} operation into the algebra $T_\sigma(V)$ of terms over a given signature $\sigma$ and set of variables $V$. For this algebra, we prove the expected universal property:
\preparalink{https://archive.softwareheritage.org/swh:1:cnt:267b5bdd29a0bef5905d5289e52708282d8e8323;origin=https://github.com/amato-gianluca/UniMath;visit=swh:1:snp:c840c96e2c4f3571fd3082b82697cbaaac480f2c;anchor=swh:1:rel:1db4fad6810e17a5a9dcaa0563a43b566aaca252;path=/UniMath/Algebra/Universal/FreeAlgebras.v;lines=18-19}
\begin{code}
  Definition |\swhl{free\_algebra}| (σ: signature) (V: varspec σ): algebra σ :=
    @make_algebra σ (termset σ V) build_term.
\end{code}
\preparalink{https://archive.softwareheritage.org/swh:1:cnt:267b5bdd29a0bef5905d5289e52708282d8e8323;origin=https://github.com/amato-gianluca/UniMath;visit=swh:1:snp:c840c96e2c4f3571fd3082b82697cbaaac480f2c;anchor=swh:1:rel:1db4fad6810e17a5a9dcaa0563a43b566aaca252;path=/UniMath/Algebra/Universal/FreeAlgebras.v;lines=42-49}
\begin{code}
  Definition |\swhl{universalmap}| (a : algebra σ) {V: varspec σ} (α: assignment a V)
    : ∑ h: free_algebra σ V ↷ a, ∏ v: V, h _ (varterm v) = α v.
\end{code}
\preparalink{https://archive.softwareheritage.org/swh:1:cnt:267b5bdd29a0bef5905d5289e52708282d8e8323;origin=https://github.com/amato-gianluca/UniMath;visit=swh:1:snp:c840c96e2c4f3571fd3082b82697cbaaac480f2c;anchor=swh:1:rel:1db4fad6810e17a5a9dcaa0563a43b566aaca252;path=/UniMath/Algebra/Universal/FreeAlgebras.v;lines=93-108}
\begin{code}
  Definition |\swhl{iscontr\_universalmap}| (a : algebra σ) {V: varspec σ} (α: assignment a V)
    : iscontr (∑ h:free_algebra σ V ↷ a, ∏ v:V, h (varsort v) (varterm v) = α v).
\end{code}

In \preparalink{https://archive.softwareheritage.org/swh:1:cnt:48c1d665d75d631c9260f114dbcba0c22c5afe69;origin=https://github.com/amato-gianluca/UniMath;visit=swh:1:snp:c840c96e2c4f3571fd3082b82697cbaaac480f2c;anchor=swh:1:rel:1db4fad6810e17a5a9dcaa0563a43b566aaca252;path=/UniMath/Algebra/Universal/TermAlgebras.v}\lncode{|\swhl{TermAlgebras.v}|} we just consider the special case of \lncode{FreeAlgebras.v} for the empty set of variables, i.e.~for ground terms. In this case, the universal mapping property is replaced by the initiality of the ground term algebra.

\subsection{Relation to W-types}\label{sec:w}
W-types are a family of inductive types first introduced in \cite{Martin-LofPer1984Itt:} as a way to encapsulate the concept of \emph{constructive} well-ordering and transfinite induction. They can be also used to express strictly positive inductive types, as proven in \cite{AbbottNestedInductive}. The work of \cite{hugunin:LIPIcs.TYPES.2020.8} shows that interesting computational properties can be retained when doing so.
When introduced on top of a given intuitionistic type theory, they provide then a robust foundation for reasoning about inductive data types and recursion within the framework of constructive mathematics.\footnote{Their semantics is detailed in \cite{berg2015wtypes,DBLP:journals/mscs/BergM18} and \cite{GambinoHyland}.}  

\smallskip

To interpret their defining rules, one can think of a W-type as a type of rooted, well-founded trees with certain constraints for branching. The formation rule
\begin{center}
\begin{prooftree}
    \AxiomC{\lncode{A : UU}}
    \AxiomC{\lncode{B : A → UU}}
    \BinaryInfC{\lncode{W A B : UU}}
\end{prooftree}    
\end{center}
requires a type \lncode{A} for the label of the nodes and a type family \lncode{B : A → UU} for specifying arities. A node labelled with \lncode{x : A} can be thought of as having  "\lncode{B(x)} many" children. Accordingly, in order to introduce a new canonical term, one needs to specify a label \lncode{x : A} for the root node and a subtree for any term of type \lncode{B(x)} by means of a function \lncode{B(x) → W A B}. This is stated in the introduction rule.
\begin{prooftree}
    \AxiomC{\lncode{x:A}}
    \AxiomC{\lncode{f : B(x) → W A B}}
    \BinaryInfC{\lncode{sup x f : W A B}}
\end{prooftree}

The elimination rule
\begin{prooftree}
    \AxiomC{\lncode{E : W A B → UU}}
    \AxiomC{\lncode{e : ∏ x f, ( ∏ (b:B(x)), E(f b) ) → E(sup x f)}}
    \BinaryInfC{\lncode{ind: ∏ (w : W A B), E(w)}}
\end{prooftree}
tells us how to inhabit
the predicate \lncode{E : W A B → UU} for all terms of type \lncode{W A B}. Given \lncode{x:A} and \sloppy\lncode{f:B(x) → W A B}, it requires us to produce a proof \lncode{e : E(sup x f)}, that is: to prove that
the predicate holds for the canonical term specified by \lncode{x} and \lncode{f}; in doing so, one has to assume that \lncode{E} holds for any of the relevant subtrees (that is \lncode{∏ (b:B(x)), E(f b)}).

Finally, the computation rule states that the proof \lncode{ind} just obtained is judgmentally the same as the one obtained by applying \lncode{e} to the proof term for subtrees obtained by \lncode{ind}.

\begin{code}
  ind (sup a f) ≡ e a f (ƛ b. ind (f b))
\end{code}

This definition of W-types is not available in UniMath. Nevertheless, it is possible to reason internally about types which behave like W-types by means of \emph{homotopy} W-types, which are presented in details by~\cite{DBLP:conf/lics/AwodeyGS12}.

Given \lncode{A : UU} and \lncode{B : A → UU}, a corresponding homotopy W-type consists of a type together with functions encapsulating the introduction and elimination principle and satisfying the appropriate computation rule w.r.t.~the equality type. This can be expressed in UniMath as follows.

\preparalink{https://archive.softwareheritage.org/swh:1:cnt:acaaca5b36835bf40030a44a69a5f55fcfe9ab15;origin=https://github.com/amato-gianluca/UniMath;visit=swh:1:snp:c840c96e2c4f3571fd3082b82697cbaaac480f2c;anchor=swh:1:rel:1db4fad6810e17a5a9dcaa0563a43b566aaca252;path=/UniMath/Induction/W/Wtypes.v;lines=16-23}
\begin{code}
  Definition |\swhl{Wtype}| (A: UU) (B: ∏ x: A, UU): UU
    := ∑ (U: UU)
       (w_sup: ∏ (x : A) (f : B x → U), U)
       (w_ind: ∏ (E : U → UU) 
                 (e_s : ∏ (x: A) (f: B x → U) (IH: ∏ u: B x, E (f u)), E (w_sup x f))
                 (w: U), E w),
       (∏ (E : U → UU)
          (e_s : ∏ (x: A) (f: B x → U) (IH: ∏ u: B x, E (f u)), E (w_sup x f))
          (x : A) (f : B x → U)
        , w_ind E e_s (w_sup x f) = e_s x f (λ u, w_ind E e_s (f u))).
\end{code}

A classical approach, as the one employed by \cite{capretta:1999}, would be to resort to W-types to define free algebras. Since general inductive definitions are not available in our formal system, we can not do that while maintaining the computational properties we are interested in.
Nonetheless, it is still expected for our structure of a free algebra to resemble that of a W-type. We show this is indeed the case in \preparalink{https://archive.softwareheritage.org/swh:1:cnt:9ca83929ea961a32fbb5a1dde5748ae91167166f;origin=https://github.com/amato-gianluca/UniMath;visit=swh:1:snp:c840c96e2c4f3571fd3082b82697cbaaac480f2c;anchor=swh:1:rel:1db4fad6810e17a5a9dcaa0563a43b566aaca252;path=/UniMath/Algebra/Universal/WTypes.v}\lncode{|\swhl{WTypes.v}|} for any ground algebra of single-sorted signature \lncode{σ}.\footnote{The same result also holds for free algebras, since they are just ground algebras for other appropriate single-sorted signatures.} 

Our main goal is to identify a type \lncode{A : UU} and a type family \lncode{B : A → UU} with the aim of constructing a homotopy W-type
\preparalink{https://archive.softwareheritage.org/swh:1:cnt:9ca83929ea961a32fbb5a1dde5748ae91167166f;origin=https://github.com/amato-gianluca/UniMath;visit=swh:1:snp:c840c96e2c4f3571fd3082b82697cbaaac480f2c;anchor=swh:1:rel:1db4fad6810e17a5a9dcaa0563a43b566aaca252;path=/UniMath/Algebra/Universal/WTypes.v;lines=563-570}
\begin{code}
  |\swhl{groundTermAlgebraWtype}|: Wtype A B := (U ,, sup ,, ind ,, beta)
\end{code}
whose first component is judgmentally the carrier type of the ground algebra with signature \lncode{σ}. So
\preparalink{https://archive.softwareheritage.org/swh:1:cnt:9ca83929ea961a32fbb5a1dde5748ae91167166f;origin=https://github.com/amato-gianluca/UniMath;visit=swh:1:snp:c840c96e2c4f3571fd3082b82697cbaaac480f2c;anchor=swh:1:rel:1db4fad6810e17a5a9dcaa0563a43b566aaca252;path=/UniMath/Algebra/Universal/WTypes.v;lines=309}
\lncode{|\swhl{U}| := gterm σ tt}.

The idea is to identify any ground term \lncode{t: U} with an inductively defined tree of ground terms. The root is \lncode{t} itself and, if a node is a term \lncode{t': U}, its children are the components of \lncode{subterms t'}. We can label each node of this tree with its principal operation which has type \preparalink{https://archive.softwareheritage.org/swh:1:cnt:9ca83929ea961a32fbb5a1dde5748ae91167166f;origin=https://github.com/amato-gianluca/UniMath;visit=swh:1:snp:c840c96e2c4f3571fd3082b82697cbaaac480f2c;anchor=swh:1:rel:1db4fad6810e17a5a9dcaa0563a43b566aaca252;path=/UniMath/Algebra/Universal/WTypes.v;lines=307}\lncode{|\swhl{A}| := names σ}. The number of children of a node labelled by \lncode{a : A} is then the arity of the label, precisely \preparalink{https://archive.softwareheritage.org/swh:1:cnt:9ca83929ea961a32fbb5a1dde5748ae91167166f;origin=https://github.com/amato-gianluca/UniMath;visit=swh:1:snp:c840c96e2c4f3571fd3082b82697cbaaac480f2c;anchor=swh:1:rel:1db4fad6810e17a5a9dcaa0563a43b566aaca252;path=/UniMath/Algebra/Universal/WTypes.v;lines=308}\lncode{|\swhl{B(a)}| := ⟦ length (arity a) ⟧.} This reasoning motivates our choices of \lncode{A} and \lncode{B} and can be expanded to identify crucial terms to derive the definitions of \lncode{sup}, \lncode{ind} and \lncode{beta}.

As a matter of fact, we have already introduced a method to obtain a term from an operation and an appropriate listing of subterms, namely
\begin{code}
  build_gterm (nm: names σ ) (v: (term σ )⋆ (arity nm)) : gterm σ (sort nm).
\end{code}
This does the same job of the introduction term
\begin{code}
  sup : ∏ (x : A), (B x → U) → U   
\end{code}
we are trying to define, but their types are not convertible. Still, by definition of \lncode{A} and since \lncode{σ} is single sorted we have that the \lncode{build_gterm}'s type is convertible to \lncode{∏ (nm: A), (gterm σ )⋆ (arity nm) → U}. Moreover we prove equivalences
\preparalink{https://archive.softwareheritage.org/swh:1:cnt:9ca83929ea961a32fbb5a1dde5748ae91167166f;origin=https://github.com/amato-gianluca/UniMath;visit=swh:1:snp:c840c96e2c4f3571fd3082b82697cbaaac480f2c;anchor=swh:1:rel:1db4fad6810e17a5a9dcaa0563a43b566aaca252;path=/UniMath/Algebra/Universal/WTypes.v;lines=313-321}
\begin{code}
  Definition |\swhl{gtweq\_sec}| (x:A) : (gterm σ)⋆ (arity x) ≃ (B x → U).
\end{code}
\preparalink{https://archive.softwareheritage.org/swh:1:cnt:9ca83929ea961a32fbb5a1dde5748ae91167166f;origin=https://github.com/amato-gianluca/UniMath;visit=swh:1:snp:c840c96e2c4f3571fd3082b82697cbaaac480f2c;anchor=swh:1:rel:1db4fad6810e17a5a9dcaa0563a43b566aaca252;path=/UniMath/Algebra/Universal/WTypes.v;lines=332-334}
\begin{code}
  Definition |\swhl{gtweqtoU}| : (∏ x : A, (gterm σ)⋆ (arity x) → U) ≃ (∏ x : A, (B x → U) → U).
\end{code}
The application of the latter to \lncode{build_gterm} is our definition of \preparalink{https://archive.softwareheritage.org/swh:1:cnt:9ca83929ea961a32fbb5a1dde5748ae91167166f;origin=https://github.com/amato-gianluca/UniMath;visit=swh:1:snp:c840c96e2c4f3571fd3082b82697cbaaac480f2c;anchor=swh:1:rel:1db4fad6810e17a5a9dcaa0563a43b566aaca252;path=/UniMath/Algebra/Universal/WTypes.v;lines=336-340}\lncode{|\swhl{sup}|}.

In a similar manner, the elimination term
\preparalink{https://archive.softwareheritage.org/swh:1:cnt:9ca83929ea961a32fbb5a1dde5748ae91167166f;origin=https://github.com/amato-gianluca/UniMath;visit=swh:1:snp:c840c96e2c4f3571fd3082b82697cbaaac480f2c;anchor=swh:1:rel:1db4fad6810e17a5a9dcaa0563a43b566aaca252;path=/UniMath/Algebra/Universal/WTypes.v;lines=441-445}
\begin{code}
  |\swhl{ind}|: ∏ E : U → UU, ind_HP E → ∏ w : U, E w
\end{code}
is the application to
\begin{code}
  term_ind : (∏ P : ∏ s : sorts σ, gterm σ s → UU, term_ind_HP P → ∏ s t, P s t)
\end{code}
of an appropriate equivalence \preparalink{https://archive.softwareheritage.org/swh:1:cnt:9ca83929ea961a32fbb5a1dde5748ae91167166f;origin=https://github.com/amato-gianluca/UniMath;visit=swh:1:snp:c840c96e2c4f3571fd3082b82697cbaaac480f2c;anchor=swh:1:rel:1db4fad6810e17a5a9dcaa0563a43b566aaca252;path=/UniMath/Algebra/Universal/WTypes.v;lines=427-439}\lncode{|\swhl{ind\_weq}|}.
We leave out its full construction, but we mention the following lemmas.
\preparalink{https://archive.softwareheritage.org/swh:1:cnt:9ca83929ea961a32fbb5a1dde5748ae91167166f;origin=https://github.com/amato-gianluca/UniMath;visit=swh:1:snp:c840c96e2c4f3571fd3082b82697cbaaac480f2c;anchor=swh:1:rel:1db4fad6810e17a5a9dcaa0563a43b566aaca252;path=/UniMath/Algebra/Universal/WTypes.v;lines=349-352}
\begin{code}
  Definition |\swhl{lower\_predicate}| : (∏ (s: sorts σ), gterm σ s → UU) ≃ (U → UU).
\end{code}
\preparalink{https://archive.softwareheritage.org/swh:1:cnt:9ca83929ea961a32fbb5a1dde5748ae91167166f;origin=https://github.com/amato-gianluca/UniMath;visit=swh:1:snp:c840c96e2c4f3571fd3082b82697cbaaac480f2c;anchor=swh:1:rel:1db4fad6810e17a5a9dcaa0563a43b566aaca252;path=/UniMath/Algebra/Universal/WTypes.v;lines=365-372}
\begin{code}
  Theorem |\swhl{ind\_HP\_Hypo}| (nm:names σ) (v : (gterm σ)⋆ (arity nm))
    : hvec (h1map_vec P v) ≃ (∏ u : B nm, (lower_predicate P) (f u)).
\end{code}
\preparalink{https://archive.softwareheritage.org/swh:1:cnt:9ca83929ea961a32fbb5a1dde5748ae91167166f;origin=https://github.com/amato-gianluca/UniMath;visit=swh:1:snp:c840c96e2c4f3571fd3082b82697cbaaac480f2c;anchor=swh:1:rel:1db4fad6810e17a5a9dcaa0563a43b566aaca252;path=/UniMath/Algebra/Universal/WTypes.v;lines=406-418}
\begin{code}
  Theorem |\swhl{HP\_weq}| : term_ind_HP P ≃ ind_HP (lower_predicate P).
\end{code}
Here, and whenever we do not explicitly bound it as a variable, \lncode{f} is just \preparalink{https://archive.softwareheritage.org/swh:1:cnt:9ca83929ea961a32fbb5a1dde5748ae91167166f;origin=https://github.com/amato-gianluca/UniMath;visit=swh:1:snp:c840c96e2c4f3571fd3082b82697cbaaac480f2c;anchor=swh:1:rel:1db4fad6810e17a5a9dcaa0563a43b566aaca252;path=/UniMath/Algebra/Universal/WTypes.v;lines=363}\swhl{notation} for \lncode{gtweq_sec nm v}, that is the image of \lncode{v} under the first equivalence introduced above. \preparalink{https://archive.softwareheritage.org/swh:1:cnt:9ca83929ea961a32fbb5a1dde5748ae91167166f;origin=https://github.com/amato-gianluca/UniMath;visit=swh:1:snp:c840c96e2c4f3571fd3082b82697cbaaac480f2c;anchor=swh:1:rel:1db4fad6810e17a5a9dcaa0563a43b566aaca252;path=/UniMath/Algebra/Universal/WTypes.v;lines=346-347}\lncode{|\swhl{ind\_HP}|} stands for the type of \lncode{e_s} in the definition of homotopy W-type. More precisely,
\begin{code}
  Definition ind_HP (E:U → UU) : UU
    := ∏ (x : A) (f : B x → U), (∏ u : B x, E (f u)) → E (sup x f).
\end{code}

Finally, we discuss the proof of the computation path
\preparalink{https://archive.softwareheritage.org/swh:1:cnt:9ca83929ea961a32fbb5a1dde5748ae91167166f;origin=https://github.com/amato-gianluca/UniMath;visit=swh:1:snp:c840c96e2c4f3571fd3082b82697cbaaac480f2c;anchor=swh:1:rel:1db4fad6810e17a5a9dcaa0563a43b566aaca252;path=/UniMath/Algebra/Universal/WTypes.v;lines=539-559}
\begin{code}
  Definition |\swhl{beta}| : ∏ E e_s x f,
    ind E e_s (sup x f) = e_s x f (λ u, ind E e_s (f u)).
\end{code}
Here the parameters \lncode{E}, \lncode{e_s} and \lncode{f} each have a type involved in a previously proved equivalence. We first consider the corresponding result quantified over the domains of these equivalences. To be clear, instead of quantifying over \lncode{f : B x → U} we do it over \lncode{v : (gterm σ)⋆ (arity nm)} and we write \lncode{gtweq_sec nm v} in place of \lncode{f}. We do the same for \lncode{E} and \lncode{e_s}.
After \preparalink{https://archive.softwareheritage.org/swh:1:cnt:9ca83929ea961a32fbb5a1dde5748ae91167166f;origin=https://github.com/amato-gianluca/UniMath;visit=swh:1:snp:c840c96e2c4f3571fd3082b82697cbaaac480f2c;anchor=swh:1:rel:1db4fad6810e17a5a9dcaa0563a43b566aaca252;path=/UniMath/Algebra/Universal/WTypes.v;lines=454-536}\swhl{inhabiting the new type}, we can prove our goal \lncode{beta} by the well-known lemma of UniMath
\preparalink{https://archive.softwareheritage.org/swh:1:cnt:7613e17ec2ea074fe36ba9cf72120b17055aeead;origin=https://github.com/amato-gianluca/UniMath;visit=swh:1:snp:c840c96e2c4f3571fd3082b82697cbaaac480f2c;anchor=swh:1:rel:1db4fad6810e17a5a9dcaa0563a43b566aaca252;path=/UniMath/Foundations/PartD.v;lines=319-321} \lncode{|\swhl{weqonsecbase}|}.
This approach allows us to manage many technical complications depending on the fact that equivalences are not judgmentally invertible.

Now, coming to the actual proof, we want to make use of
\begin{code}
  term_ind_step (P: ∏ (s: sorts σ), term σ s → UU) (R: term_ind_HP P) 
                (nm: names σ) (v: (term σ)⋆ (arity nm))
    : term_ind P R (build_term nm v) 
      = R nm v (h2map (λ s t q, term_ind P R t) (h1lift v)).
\end{code}
which, once again, is the intended path in the wrong types. To conclude the proof, it suffices to find an equivalence which maps (propositionally) both sides of \lncode{term_ind_step} to the corresponding sides of \lncode{beta}. This is a delicate step, since the equivalence we choose here is proof relevant: its proof term must be manageable and it must interact nicely with many of the other constructions presented until now. We opt for
\preparalink{https://archive.softwareheritage.org/swh:1:cnt:9ca83929ea961a32fbb5a1dde5748ae91167166f;origin=https://github.com/amato-gianluca/UniMath;visit=swh:1:snp:c840c96e2c4f3571fd3082b82697cbaaac480f2c;anchor=swh:1:rel:1db4fad6810e17a5a9dcaa0563a43b566aaca252;path=/UniMath/Algebra/Universal/WTypes.v;lines=374-402}
\begin{code}
  Theorem |\swhl{ind\_HP\_Th}| (nm:names σ) (v : (gterm σ)⋆ (arity nm))
    : P (sort nm) (build_gterm nm v) ≃ (lower_predicate P) (sup nm f).
\end{code}
which is actually a lemma we already used to construct \lncode{HP_weq}.

Proving that this equivalence respects the right hand sides of \lncode{term_ind_step} and \lncode{beta} is not trivial. In particular, the proof of theorem \preparalink{https://archive.softwareheritage.org/swh:1:cnt:9ca83929ea961a32fbb5a1dde5748ae91167166f;origin=https://github.com/amato-gianluca/UniMath;visit=swh:1:snp:c840c96e2c4f3571fd3082b82697cbaaac480f2c;anchor=swh:1:rel:1db4fad6810e17a5a9dcaa0563a43b566aaca252;path=/UniMath/Algebra/Universal/WTypes.v;lines=505-516}\lncode{|\swhl{ind\_HP\_Hypo\_h2map}|} revealed a rather challenging task, because of several technicalities in relating our heterogeneous vectors to dependent functions.\footnote{A possible refinement and revision of the \lncode{HVectors.v} module might simplify some of the subtleties involved in the proof of this central theorem.}

\subsection{Equations and equational algebras}\label{sec:eq}

Equations and their associated structures are key notions in universal algebra.
Although an extensive treatment of equational algebras and varieties is out of the scope of the present work, the basic definitions are already present in our implementation in the file \preparalink{https://archive.softwareheritage.org/swh:1:cnt:4477e7c93eaf7aed2b05b91cb5ac08f96e754718;origin=https://github.com/amato-gianluca/UniMath;visit=swh:1:snp:c840c96e2c4f3571fd3082b82697cbaaac480f2c;anchor=swh:1:rel:1db4fad6810e17a5a9dcaa0563a43b566aaca252;path=/UniMath/Algebra/Universal/EqAlgebras.v}\lncode{|\swhl{EqAlgebras.v}|}.

In our setting, an {equation} is a pair of terms (with variables) of the same sort.  Their intended meaning is to specify {identities law} where variables are implicitly universally quantified.
\preparalink{https://archive.softwareheritage.org/swh:1:cnt:e759d6f54ecfb5252e1e288cfd605bb7a0fd35c9;origin=https://github.com/amato-gianluca/UniMath;visit=swh:1:snp:c840c96e2c4f3571fd3082b82697cbaaac480f2c;anchor=swh:1:rel:1db4fad6810e17a5a9dcaa0563a43b566aaca252;path=/UniMath/Algebra/Universal/Equations.v;lines=15-16}
\begin{code}
  Definition |\swhl{equation}| (σ : signature) (V: varspec σ): UU
    := ∑ s: sorts σ, term σ V s × term σ V s.
\end{code}
The associated projections are denoted \preparalink{https://archive.softwareheritage.org/swh:1:cnt:e759d6f54ecfb5252e1e288cfd605bb7a0fd35c9;origin=https://github.com/amato-gianluca/UniMath;visit=swh:1:snp:c840c96e2c4f3571fd3082b82697cbaaac480f2c;anchor=swh:1:rel:1db4fad6810e17a5a9dcaa0563a43b566aaca252;path=/UniMath/Algebra/Universal/Equations.v;lines=18-19}\lncode{|\swhl{eqsort}|}, \preparalink{https://archive.softwareheritage.org/swh:1:cnt:e759d6f54ecfb5252e1e288cfd605bb7a0fd35c9;origin=https://github.com/amato-gianluca/UniMath;visit=swh:1:snp:c840c96e2c4f3571fd3082b82697cbaaac480f2c;anchor=swh:1:rel:1db4fad6810e17a5a9dcaa0563a43b566aaca252;path=/UniMath/Algebra/Universal/Equations.v;lines=21}\lncode{|\swhl{lhs}|}, and \preparalink{https://archive.softwareheritage.org/swh:1:cnt:e759d6f54ecfb5252e1e288cfd605bb7a0fd35c9;origin=https://github.com/amato-gianluca/UniMath;visit=swh:1:snp:c840c96e2c4f3571fd3082b82697cbaaac480f2c;anchor=swh:1:rel:1db4fad6810e17a5a9dcaa0563a43b566aaca252;path=/UniMath/Algebra/Universal/Equations.v;lines=23}\lncode{|\swhl{rhs}|} respectively.
An {equation system} is just a family of equations.
\preparalink{https://archive.softwareheritage.org/swh:1:cnt:e759d6f54ecfb5252e1e288cfd605bb7a0fd35c9;origin=https://github.com/amato-gianluca/UniMath;visit=swh:1:snp:c840c96e2c4f3571fd3082b82697cbaaac480f2c;anchor=swh:1:rel:1db4fad6810e17a5a9dcaa0563a43b566aaca252;path=/UniMath/Algebra/Universal/Equations.v;lines=43-44}
\begin{code}
  Definition |\swhl{eqsystem}| (σ : signature) (V: varspec σ): UU
    := ∑ E : UU, E → equation σ V.
\end{code}
Then, we pack all the above data into an {equational specification}, that is a signature endowed with an equation system (and the necessary variable specification).
\preparalink{https://archive.softwareheritage.org/swh:1:cnt:e759d6f54ecfb5252e1e288cfd605bb7a0fd35c9;origin=https://github.com/amato-gianluca/UniMath;visit=swh:1:snp:c840c96e2c4f3571fd3082b82697cbaaac480f2c;anchor=swh:1:rel:1db4fad6810e17a5a9dcaa0563a43b566aaca252;path=/UniMath/Algebra/Universal/Equations.v;lines=58}
\begin{code}
  Definition |\swhl{eqspec}|: UU  := ∑ (σ : signature) (V: varspec σ), eqsystem σ V.
\end{code}

The interpretation of an equation is easily defined using the \preparalink{https://archive.softwareheritage.org/swh:1:cnt:485d80e1be15fce12cfba0244b95752fd1bf4bb3;origin=https://github.com/amato-gianluca/UniMath;visit=swh:1:snp:c840c96e2c4f3571fd3082b82697cbaaac480f2c;anchor=swh:1:rel:1db4fad6810e17a5a9dcaa0563a43b566aaca252;path=/UniMath/Algebra/Universal/VTerms.v;lines=51-60}\swhl{general version} (admitting variables) of the function \lncode{fromterm} introduced in Sect.~\ref{sec:induction}.
More precisely, the predicate \preparalink{https://archive.softwareheritage.org/swh:1:cnt:4477e7c93eaf7aed2b05b91cb5ac08f96e754718;origin=https://github.com/amato-gianluca/UniMath;visit=swh:1:snp:c840c96e2c4f3571fd3082b82697cbaaac480f2c;anchor=swh:1:rel:1db4fad6810e17a5a9dcaa0563a43b566aaca252;path=/UniMath/Algebra/Universal/EqAlgebras.v;lines=15-17}\lncode{|\swhl{holds}|} that checks if the universal closure of an equation \lncode{e} holds in an algebra is given as follows:
\begin{code}
  Definition holds {σ: signature} {V: varspec σ}
                   (a: algebra σ) (e: equation σ V) : UU
    := ∏ α, fromterm (ops a) α (eqsort e) (lhs e) = fromterm (ops a) α (eqsort e) (rhs e).
\end{code}
From this, it is immediate to define the type \preparalink{https://archive.softwareheritage.org/swh:1:cnt:4477e7c93eaf7aed2b05b91cb5ac08f96e754718;origin=https://github.com/amato-gianluca/UniMath;visit=swh:1:snp:c840c96e2c4f3571fd3082b82697cbaaac480f2c;anchor=swh:1:rel:1db4fad6810e17a5a9dcaa0563a43b566aaca252;path=/UniMath/Algebra/Universal/EqAlgebras.v;lines=19-23}\lncode{|\swhl{eqalgebra}|} of {equational algebras} as those algebras in which all the equations of a given equational specification hold.

\subsection{Categorical structures}\label{sec:cat}

Universal algebra has a natural and fruitful interplay with category theory, as discussed by, e.g., \cite{hylandpower}.
As claimed in the introduction, our mechanisation includes basic categorical constructions for organizing and reasoning about universal algebra structures. 
In agreement with the general philosophy of univalent mathematics,\footnote{See the remarks in~\cite{10.1007/978-3-319-21284-5_14}, where category theory was introduced first in a HoTT-setting.} we can prove that the categories we are interested in -- of algebras and equational algebras -- are univalent indeed.

In order to develop formal proofs of that property, two possible strategies are available.

\noindent A simplest one consists of building the desired category from scratch, and then prove that univalence holds between any pair of isomorphic objects. However, experience has shown that this strategy often lacks a certain naturalness, and it makes the steps involved in the construction hard.

The second available strategy has revealed practicable in a more efficient way: we define the desired category in a step-by-step construction by adding {layers} to a {base category} already given. Such a notion of layer corresponds precisely to a \emph{displayed category} as formulated by~\cite{lmcs:5252}. Displayed categories can be thought of as the type-theoretic counterpart of fibrations, and constitute a widely adopted instrument to reason about categories even at higher dimensions in the UniMath library.\footnote{See e.g.~\citep{ahrens_et_al:LIPIcs:2019:10512}.}

To this end, a simple approach would be to proceed in two separate steps, first build the desired categories, then write the proofs that they are univalent.
After defining a displayed category over a base category, we can then build a {total category} whose univalence is proven by checking univalence for the base category \emph{and} a displayed version of univalence for the category displayed over the base. This is a generalised version of the so-called \emph{structure identity principle}, introduced first by~\cite{aczel2011voevodsky} as invariance of all structural properties of isomorphic structures (broadly considered).

Since the type of morphisms in UniMath's categories are sets, we need to restrict our attention to algebras whose carrier is not just and index type but an indexed hSet (denoted as \preparalink{https://archive.softwareheritage.org/swh:1:cnt:d66adf476af0590fc10873027c24c742b1160125;origin=https://github.com/amato-gianluca/UniMath;visit=swh:1:snp:c840c96e2c4f3571fd3082b82697cbaaac480f2c;anchor=swh:1:rel:1db4fad6810e17a5a9dcaa0563a43b566aaca252;path=/UniMath/Algebra/Universal/SortedTypes.v;lines=56}\lncode{|\swhl{shSet}|} in the library). The special case of algebras whose support is an \lncode{shSet} is the \preparalink{https://archive.softwareheritage.org/swh:1:cnt:3a898cb11b366ae4c72e95545d09cdadcc20534a;origin=https://github.com/amato-gianluca/UniMath;visit=swh:1:snp:c840c96e2c4f3571fd3082b82697cbaaac480f2c;anchor=swh:1:rel:1db4fad6810e17a5a9dcaa0563a43b566aaca252;path=/UniMath/Algebra/Universal/Algebras.v;lines=228-229}\lncode{|\swhl{hSetAlgebra}|} type.

To build our category of algebras, we apply that very principle: the structure of algebras and homomorphisms is displayed over a base category of sorted hSets, as proven in our \preparalink{https://archive.softwareheritage.org/swh:1:cnt:633f4e825754b7d1a89bd0cb4997e20c565dc1c5;origin=https://github.com/amato-gianluca/UniMath;visit=swh:1:snp:c840c96e2c4f3571fd3082b82697cbaaac480f2c;anchor=swh:1:rel:1db4fad6810e17a5a9dcaa0563a43b566aaca252;path=/UniMath/CategoryTheory/categories/Universal_Algebra/Algebras.v;lines=64-76}\swhl{lemma}.

In a bit more detailed manner, when building the main category of algebras over a given signature $\sigma$,
\begin{itemize}
\item We \preparalink{https://archive.softwareheritage.org/swh:1:cnt:633f4e825754b7d1a89bd0cb4997e20c565dc1c5;origin=https://github.com/amato-gianluca/UniMath;visit=swh:1:snp:c840c96e2c4f3571fd3082b82697cbaaac480f2c;anchor=swh:1:rel:1db4fad6810e17a5a9dcaa0563a43b566aaca252;path=/UniMath/CategoryTheory/categories/Universal_Algebra/Algebras.v;lines=67}\swhl{associate} to each sorted-hSet its family of algebras;
\item To each sorted-function, we \preparalink{https://archive.softwareheritage.org/swh:1:cnt:633f4e825754b7d1a89bd0cb4997e20c565dc1c5;origin=https://github.com/amato-gianluca/UniMath;visit=swh:1:snp:c840c96e2c4f3571fd3082b82697cbaaac480f2c;anchor=swh:1:rel:1db4fad6810e17a5a9dcaa0563a43b566aaca252;path=/UniMath/CategoryTheory/categories/Universal_Algebra/Algebras.v;lines=68}\swhl{associate} the property \lncode{ishom};
\item We then use the \preparalink{https://archive.softwareheritage.org/swh:1:cnt:633f4e825754b7d1a89bd0cb4997e20c565dc1c5;origin=https://github.com/amato-gianluca/UniMath;visit=swh:1:snp:c840c96e2c4f3571fd3082b82697cbaaac480f2c;anchor=swh:1:rel:1db4fad6810e17a5a9dcaa0563a43b566aaca252;path=/UniMath/CategoryTheory/categories/Universal_Algebra/Algebras.v;lines=73-75}\swhl{fact} that the identity sorted-function defines an algebra homomorphism, and that \lncode{ishom} is closed under composition of sorted-functions, as stated by \lncode{ishomid} and \lncode{ishomcomp}, respectively;\footnote{See the end of Section \ref{sec:alg}.}
\item Finally, we use the UniMath lemma \preparalink{https://archive.softwareheritage.org/swh:1:cnt:87b830ce1c21521ef5a60529eeb4ac604718a17e;origin=https://github.com/amato-gianluca/UniMath;visit=swh:1:snp:c840c96e2c4f3571fd3082b82697cbaaac480f2c;anchor=swh:1:rel:1db4fad6810e17a5a9dcaa0563a43b566aaca252;path=/UniMath/CategoryTheory/DisplayedCats/SIP.v;lines=47-59}\lncode{|\swhl{is\_univalent\_disp\_from\_SIP\_data}|} to \preparalink{https://archive.softwareheritage.org/swh:1:cnt:633f4e825754b7d1a89bd0cb4997e20c565dc1c5;origin=https://github.com/amato-gianluca/UniMath;visit=swh:1:snp:c840c96e2c4f3571fd3082b82697cbaaac480f2c;anchor=swh:1:rel:1db4fad6810e17a5a9dcaa0563a43b566aaca252;path=/UniMath/CategoryTheory/categories/Universal_Algebra/Algebras.v;lines=78-86}\swhl{prove displayed univalence} by showing that the property of being an algebra is an hSet indeed, and that any two interpretations of symbols of $\sigma$ are equal whenever the identity sorted-function is an homomorphism w.r.t.~these given assignments.
\end{itemize}

At this point, proving that the base category of shSets \preparalink{https://archive.softwareheritage.org/swh:1:cnt:633f4e825754b7d1a89bd0cb4997e20c565dc1c5;origin=https://github.com/amato-gianluca/UniMath;visit=swh:1:snp:c840c96e2c4f3571fd3082b82697cbaaac480f2c;anchor=swh:1:rel:1db4fad6810e17a5a9dcaa0563a43b566aaca252;path=/UniMath/CategoryTheory/categories/Universal_Algebra/Algebras.v;lines=90-167}\swhl{is univalent} revealed already non-trivial. Nevertheless, we managed on the issue by tweaking the proof-terms already constructed for functor categories in UniMath.
The resulting total category of algebras \preparalink{https://archive.softwareheritage.org/swh:1:cnt:633f4e825754b7d1a89bd0cb4997e20c565dc1c5;origin=https://github.com/amato-gianluca/UniMath;visit=swh:1:snp:c840c96e2c4f3571fd3082b82697cbaaac480f2c;anchor=swh:1:rel:1db4fad6810e17a5a9dcaa0563a43b566aaca252;path=/UniMath/CategoryTheory/categories/Universal_Algebra/Algebras.v;lines=171-174}\swhl{is therefore univalent} in the usual sense.

\vspace{.28cm}

Turning now to equational algebras, we do not have to start the construction again from scratch: within the displayed category formalism we can identify the ``substructure'' of algebras over shSets satisfying a system of equations. In other terms, we can \preparalink{https://archive.softwareheritage.org/swh:1:cnt:50ca81bb47afef13a535f00f95b62cfa6e23a388;origin=https://github.com/amato-gianluca/UniMath;visit=swh:1:snp:c840c96e2c4f3571fd3082b82697cbaaac480f2c;anchor=swh:1:rel:1db4fad6810e17a5a9dcaa0563a43b566aaca252;path=/UniMath/CategoryTheory/categories/Universal_Algebra/EqAlgebras.v;lines=37}\swhl{take} for equational algebras the layer over the category of shSets made of the {full displayed subcategory} of the displayed category of algebras identified by the type \preparalink{https://archive.softwareheritage.org/swh:1:cnt:4477e7c93eaf7aed2b05b91cb5ac08f96e754718;origin=https://github.com/amato-gianluca/UniMath;visit=swh:1:snp:c840c96e2c4f3571fd3082b82697cbaaac480f2c;anchor=swh:1:rel:1db4fad6810e17a5a9dcaa0563a43b566aaca252;path=/UniMath/Algebra/Universal/EqAlgebras.v;lines=19-20}\lncode{|\swhl{is\_eqalgebra}|}.

Again, \preparalink{https://archive.softwareheritage.org/swh:1:cnt:50ca81bb47afef13a535f00f95b62cfa6e23a388;origin=https://github.com/amato-gianluca/UniMath;visit=swh:1:snp:c840c96e2c4f3571fd3082b82697cbaaac480f2c;anchor=swh:1:rel:1db4fad6810e17a5a9dcaa0563a43b566aaca252;path=/UniMath/CategoryTheory/categories/Universal_Algebra/EqAlgebras.v;lines=26-35}\swhl{proving} displayed univalence for this layer is not difficult, so that the total category of equational algebras over a system of equations is univalent, as required.

Finally, we rephrase the universal property of the term algebra shown in Section \ref{sec:term}: we can state its initiality in the category of algebras over a given $\sigma$ by means of the \preparalink{https://archive.softwareheritage.org/swh:1:cnt:633f4e825754b7d1a89bd0cb4997e20c565dc1c5;origin=https://github.com/amato-gianluca/UniMath;visit=swh:1:snp:c840c96e2c4f3571fd3082b82697cbaaac480f2c;anchor=swh:1:rel:1db4fad6810e17a5a9dcaa0563a43b566aaca252;path=/UniMath/CategoryTheory/categories/Universal_Algebra/Algebras.v;lines=176-182}\swhl{proof-term} made of the of the algebra itself and the contractibility of out-going homomorphisms, previously constructed.

The reader interested in the details of these categorical results is referred to our code located in the subdirectory \preparalink{https://archive.softwareheritage.org/swh:1:dir:3c79d052bbeab5f408c5966b8e948be1f74fe07f;origin=https://github.com/amato-gianluca/UniMath;visit=swh:1:snp:c840c96e2c4f3571fd3082b82697cbaaac480f2c;anchor=swh:1:rel:1db4fad6810e17a5a9dcaa0563a43b566aaca252;path=/UniMath/CategoryTheory/categories/Universal_Algebra/}\lncode{|\swhl{Universal\_Algebra}|}.

\section{Three applications}\label{sec:ex}

In this section, we want to illustrate by simple examples how to use our framework in three different settings.

\subsection{List algebras}\label{sec:lists}

We start with a very simple multi-sorted example.
We will show how to specify a signature in our framework and how to interpret a list datatype as an algebra.\footnote{The code for this example can be found in the module \preparalink{https://archive.softwareheritage.org/swh:1:cnt:03004ef64a272d40cd17833fc651273f72011f03;origin=https://github.com/amato-gianluca/UniMath;visit=swh:1:snp:c840c96e2c4f3571fd3082b82697cbaaac480f2c;anchor=swh:1:rel:1db4fad6810e17a5a9dcaa0563a43b566aaca252;path=/UniMath/Algebra/Universal/Examples/ListDataType.v}\swhl{\texttt{ListDataType.v}}.}

\vspace{.28cm}

We will need two sorts, one for elements and the other for lists.
Correspondingly, we name the two elements \preparalink{https://archive.softwareheritage.org/swh:1:cnt:03004ef64a272d40cd17833fc651273f72011f03;origin=https://github.com/amato-gianluca/UniMath;visit=swh:1:snp:c840c96e2c4f3571fd3082b82697cbaaac480f2c;anchor=swh:1:rel:1db4fad6810e17a5a9dcaa0563a43b566aaca252;path=/UniMath/Algebra/Universal/Examples/ListDataType.v;lines=15}\lncode{|\swhl{●0}|} and \preparalink{https://archive.softwareheritage.org/swh:1:cnt:03004ef64a272d40cd17833fc651273f72011f03;origin=https://github.com/amato-gianluca/UniMath;visit=swh:1:snp:c840c96e2c4f3571fd3082b82697cbaaac480f2c;anchor=swh:1:rel:1db4fad6810e17a5a9dcaa0563a43b566aaca252;path=/UniMath/Algebra/Universal/Examples/ListDataType.v;lines=16}\lncode{|\swhl{●1}|} of the standard finite set with two elements \lncode{⟦ 2 ⟧}.
\begin{code}
  Definition elem_sort_idx: ⟦ 2 ⟧ := ●0.
  Definition list_sort_idx: ⟦ 2 ⟧ := ●1.
\end{code}
Our signature for the language of lists will consist of two operation symbols for the usual constructors \emph{nil} and \emph{cons} respectively.

Such a signature is encoded with a list of pairs.  Each pair describe the input (a list of sorts) and the output (a sort) for the corresponding constructor.
\preparalink{https://archive.softwareheritage.org/swh:1:cnt:03004ef64a272d40cd17833fc651273f72011f03;origin=https://github.com/amato-gianluca/UniMath;visit=swh:1:snp:c840c96e2c4f3571fd3082b82697cbaaac480f2c;anchor=swh:1:rel:1db4fad6810e17a5a9dcaa0563a43b566aaca252;path=/UniMath/Algebra/Universal/Examples/ListDataType.v;lines=18-24}
\begin{code}
  Definition |\swhl{list\_signature}|: signature_simple
    := make_signature_simple
         [ ( nil ,, list_sort_idx ) ;
           ( [elem_sort_idx ; list_sort_idx] ,, list_sort_idx ) ]
\end{code}
For enhanced readability, we assign explicit names to the operator symbols.
\preparalink{https://archive.softwareheritage.org/swh:1:cnt:03004ef64a272d40cd17833fc651273f72011f03;origin=https://github.com/amato-gianluca/UniMath;visit=swh:1:snp:c840c96e2c4f3571fd3082b82697cbaaac480f2c;anchor=swh:1:rel:1db4fad6810e17a5a9dcaa0563a43b566aaca252;path=/UniMath/Algebra/Universal/Examples/ListDataType.v;lines=28}
\begin{code}
  Definition |\swhl{nil\_idx}|: names list_signature := ●0.
\end{code}
\preparalink{https://archive.softwareheritage.org/swh:1:cnt:03004ef64a272d40cd17833fc651273f72011f03;origin=https://github.com/amato-gianluca/UniMath;visit=swh:1:snp:c840c96e2c4f3571fd3082b82697cbaaac480f2c;anchor=swh:1:rel:1db4fad6810e17a5a9dcaa0563a43b566aaca252;path=/UniMath/Algebra/Universal/Examples/ListDataType.v;lines=29}
\begin{code}
  Definition |\swhl{cons\_idx}|: names list_signature := ●1.
\end{code}
Now, we can endow the \preparalink{https://archive.softwareheritage.org/swh:1:cnt:3750de471510aa9d8e821ce739d449c1cc8a78b5;origin=https://github.com/amato-gianluca/UniMath;visit=swh:1:snp:c840c96e2c4f3571fd3082b82697cbaaac480f2c;anchor=swh:1:rel:1db4fad6810e17a5a9dcaa0563a43b566aaca252;path=/UniMath/Combinatorics/Lists.v;lines=21}\lncode{|\swhl{list}|} datatype with the structure of an algebra over \lncode{list_signature} by using the list constructors \preparalink{https://archive.softwareheritage.org/swh:1:cnt:3750de471510aa9d8e821ce739d449c1cc8a78b5;origin=https://github.com/amato-gianluca/UniMath;visit=swh:1:snp:c840c96e2c4f3571fd3082b82697cbaaac480f2c;anchor=swh:1:rel:1db4fad6810e17a5a9dcaa0563a43b566aaca252;path=/UniMath/Combinatorics/Lists.v;lines=30-31}\lncode{|\swhl{nil}|} and \preparalink{https://archive.softwareheritage.org/swh:1:cnt:3750de471510aa9d8e821ce739d449c1cc8a78b5;origin=https://github.com/amato-gianluca/UniMath;visit=swh:1:snp:c840c96e2c4f3571fd3082b82697cbaaac480f2c;anchor=swh:1:rel:1db4fad6810e17a5a9dcaa0563a43b566aaca252;path=/UniMath/Combinatorics/Lists.v;lines=33-35}\lncode{|\swhl{cons}|}.

We fix a type \lncode{A} for our elements.
Then, the class of algebras over \lncode{list_signature} is given by\footnote{Since working with functions of type \lncode{A⋆ (arity nm) → A (sort nm)} is cumbersome, we \preparalink{https://archive.softwareheritage.org/swh:1:cnt:3a898cb11b366ae4c72e95545d09cdadcc20534a;origin=https://github.com/amato-gianluca/UniMath;visit=swh:1:snp:c840c96e2c4f3571fd3082b82697cbaaac480f2c;anchor=swh:1:rel:1db4fad6810e17a5a9dcaa0563a43b566aaca252;path=/UniMath/Algebra/Universal/Algebras.v;lines=201-224}\swhl{have included} primed version of the algebra constructors which take curried functions of type \lncode{A v0 → A v1 → ... → A vn → A s} and convert them automatically to  functions \lncode{A⋆ [v0; v1; ...; vn] → A s}, significantly simplifying the definition of new algebras.}
\preparalink{https://archive.softwareheritage.org/swh:1:cnt:03004ef64a272d40cd17833fc651273f72011f03;origin=https://github.com/amato-gianluca/UniMath;visit=swh:1:snp:c840c96e2c4f3571fd3082b82697cbaaac480f2c;anchor=swh:1:rel:1db4fad6810e17a5a9dcaa0563a43b566aaca252;path=/UniMath/Algebra/Universal/Examples/ListDataType.v;lines=31-35}
\begin{code}
  Definition |\swhl{list\_algebra}| : algebra list_signature
    := make_algebra_simple' list_signature
        [( A ; list A )]
        [( nil ; cons )].
\end{code}

From now on in this section, lemmas are just simple verification of convertibility.
They are all proven by reflexivity and the proof scripts are omitted.

To begin with, we check that the sort of elements is $A$ and the sort of lists is given by the associated list datatype:
\preparalink{https://archive.softwareheritage.org/swh:1:cnt:03004ef64a272d40cd17833fc651273f72011f03;origin=https://github.com/amato-gianluca/UniMath;visit=swh:1:snp:c840c96e2c4f3571fd3082b82697cbaaac480f2c;anchor=swh:1:rel:1db4fad6810e17a5a9dcaa0563a43b566aaca252;path=/UniMath/Algebra/Universal/Examples/ListDataType.v;lines=40-43}
\begin{code}
  Lemma |\swhl{elem\_sort\_id}| : supportset list_algebra elem_sort_idx = A.
\end{code}
\preparalink{https://archive.softwareheritage.org/swh:1:cnt:03004ef64a272d40cd17833fc651273f72011f03;origin=https://github.com/amato-gianluca/UniMath;visit=swh:1:snp:c840c96e2c4f3571fd3082b82697cbaaac480f2c;anchor=swh:1:rel:1db4fad6810e17a5a9dcaa0563a43b566aaca252;path=/UniMath/Algebra/Universal/Examples/ListDataType.v;lines=45-48}
\begin{code}
  Lemma |\swhl{list\_sort\_id}| : supportset list_algebra list_sort_idx = list A.
\end{code}
Next, we inspect the associated algebra operations. We can extract and currify them with \preparalink{https://archive.softwareheritage.org/swh:1:cnt:3a898cb11b366ae4c72e95545d09cdadcc20534a;origin=https://github.com/amato-gianluca/UniMath;visit=swh:1:snp:c840c96e2c4f3571fd3082b82697cbaaac480f2c;anchor=swh:1:rel:1db4fad6810e17a5a9dcaa0563a43b566aaca252;path=/UniMath/Algebra/Universal/Algebras.v;lines=199}\lncode{|\swhl{ops'}|}.
\noindent First, let's consider the empty list constructor.
\preparalink{https://archive.softwareheritage.org/swh:1:cnt:03004ef64a272d40cd17833fc651273f72011f03;origin=https://github.com/amato-gianluca/UniMath;visit=swh:1:snp:c840c96e2c4f3571fd3082b82697cbaaac480f2c;anchor=swh:1:rel:1db4fad6810e17a5a9dcaa0563a43b566aaca252;path=/UniMath/Algebra/Universal/Examples/ListDataType.v;lines=50}
\begin{code}
  Definition |\swhl{list\_nil}| : listset A := ops' list_algebra nil_idx tt.
\end{code}
As expected, it reduces to the usual \emph{nil} constructor.
\preparalink{https://archive.softwareheritage.org/swh:1:cnt:03004ef64a272d40cd17833fc651273f72011f03;origin=https://github.com/amato-gianluca/UniMath;visit=swh:1:snp:c840c96e2c4f3571fd3082b82697cbaaac480f2c;anchor=swh:1:rel:1db4fad6810e17a5a9dcaa0563a43b566aaca252;path=/UniMath/Algebra/Universal/Examples/ListDataType.v;lines=52-55}
\begin{code}
  Lemma |\swhl{list\_nil\_id}| : list_nil = @nil A.
\end{code}
For the list \emph{cons} constructor, the situation is more complicated.  The domain of the constructor is the product \lncode|A × listset A × unit|, meaning that the constructor has two (uncurried) arguments
\preparalink{https://archive.softwareheritage.org/swh:1:cnt:03004ef64a272d40cd17833fc651273f72011f03;origin=https://github.com/amato-gianluca/UniMath;visit=swh:1:snp:c840c96e2c4f3571fd3082b82697cbaaac480f2c;anchor=swh:1:rel:1db4fad6810e17a5a9dcaa0563a43b566aaca252;path=/UniMath/Algebra/Universal/Examples/ListDataType.v;lines=57-61}
\begin{code}
  Lemma |\swhl{list\_cons\_dom\_id}| : dom list_algebra cons_idx = A × listset A × unit.
\end{code}
Still, it reduces to the usual list cons.

\preparalink{https://archive.softwareheritage.org/swh:1:cnt:03004ef64a272d40cd17833fc651273f72011f03;origin=https://github.com/amato-gianluca/UniMath;visit=swh:1:snp:c840c96e2c4f3571fd3082b82697cbaaac480f2c;anchor=swh:1:rel:1db4fad6810e17a5a9dcaa0563a43b566aaca252;path=/UniMath/Algebra/Universal/Examples/ListDataType.v;lines=63-64}
\begin{code}
  Definition |\swhl{list\_cons}| (A: hSet) : A → listset A → listset A
    := ops' (list_algebra A) cons_idx.
\end{code}
\preparalink{https://archive.softwareheritage.org/swh:1:cnt:03004ef64a272d40cd17833fc651273f72011f03;origin=https://github.com/amato-gianluca/UniMath;visit=swh:1:snp:c840c96e2c4f3571fd3082b82697cbaaac480f2c;anchor=swh:1:rel:1db4fad6810e17a5a9dcaa0563a43b566aaca252;path=/UniMath/Algebra/Universal/Examples/ListDataType.v;lines=66-69}
\begin{code}
  Lemma |\swhl{cons\_nil\_id}| : list_cons = @cons.
\end{code}

\subsection{Equational algebras of monoids}\label{sec:grp}

From now on, we will consider single sorted examples for the sake of simplicity.

In this Section, we will discuss the \lncode{eqalgebra} of monoids.\footnote{The code for this example can be found in the module \preparalink{https://archive.softwareheritage.org/swh:1:cnt:3a5262a9e829e569e458851f429af03dee73e634;origin=https://github.com/amato-gianluca/UniMath;visit=swh:1:snp:c840c96e2c4f3571fd3082b82697cbaaac480f2c;anchor=swh:1:rel:1db4fad6810e17a5a9dcaa0563a43b566aaca252;path=/UniMath/Algebra/Universal/Examples/Monoid.v}\swhl{\texttt{Monoid.v}}.}

\vspace{.28cm}

To define single sorted signatures, our function \lncode{make_signature_simple_single_sorted} is a handy shorthand -- introduced in Section \ref{sec:alg} -- taking only a list of natural numbers.
\preparalink{https://archive.softwareheritage.org/swh:1:cnt:3a5262a9e829e569e458851f429af03dee73e634;origin=https://github.com/amato-gianluca/UniMath;visit=swh:1:snp:c840c96e2c4f3571fd3082b82697cbaaac480f2c;anchor=swh:1:rel:1db4fad6810e17a5a9dcaa0563a43b566aaca252;path=/UniMath/Algebra/Universal/Examples/Monoid.v;lines=19}
\begin{code}
  Definition |\swhl{monoid\_signature}| := make_signature_simple_single_sorted [2; 0].
\end{code}
Monoids are already defined in UniMath: given \lncode{M} a \preparalink{https://archive.softwareheritage.org/swh:1:cnt:51ab805f9977b358fc9de10d20558015c90ad566;origin=https://github.com/amato-gianluca/UniMath;visit=swh:1:snp:c840c96e2c4f3571fd3082b82697cbaaac480f2c;anchor=swh:1:rel:1db4fad6810e17a5a9dcaa0563a43b566aaca252;path=/UniMath/Algebra/Monoids.v;lines=55}\lncode{|\swhl{monoid}|}, \preparalink{https://archive.softwareheritage.org/swh:1:cnt:51ab805f9977b358fc9de10d20558015c90ad566;origin=https://github.com/amato-gianluca/UniMath;visit=swh:1:snp:c840c96e2c4f3571fd3082b82697cbaaac480f2c;anchor=swh:1:rel:1db4fad6810e17a5a9dcaa0563a43b566aaca252;path=/UniMath/Algebra/Monoids.v;lines=66}\lncode{|\swhl{unel}| M} accesses its identity and \preparalink{https://archive.softwareheritage.org/swh:1:cnt:9ccc9ebc41af334dfa8cbbbbaccbe13451111bd6;origin=https://github.com/amato-gianluca/UniMath;visit=swh:1:snp:c840c96e2c4f3571fd3082b82697cbaaac480f2c;anchor=swh:1:rel:1db4fad6810e17a5a9dcaa0563a43b566aaca252;path=/UniMath/Algebra/BinaryOperations.v;lines=1537}\lncode{|\swhl{op}|} accesses its operation.

Similarly to what we did in the previous section with lists, we endow monoids with the structure of a monoid algebra.
\preparalink{https://archive.softwareheritage.org/swh:1:cnt:3a5262a9e829e569e458851f429af03dee73e634;origin=https://github.com/amato-gianluca/UniMath;visit=swh:1:snp:c840c96e2c4f3571fd3082b82697cbaaac480f2c;anchor=swh:1:rel:1db4fad6810e17a5a9dcaa0563a43b566aaca252;path=/UniMath/Algebra/Universal/Examples/Monoid.v;lines=25-27}
\begin{code}
  Definition |\swhl{monoid\_algebra}| (M: monoid): algebra monoid_signature
    := make_algebra_simple_single_sorted' monoid_signature M
        [( op ; unel M )].
\end{code}
Next, we provide a variable specification, i.e.~an hSet of variables together with a map from variables to sorts. Since \lncode{monoid_signature} is single-sorted, the only available sort is \lncode{tt}. 

Then,  we build the associated algebra of open terms\footnote{Here \preparalink{https://archive.softwareheritage.org/swh:1:cnt:485d80e1be15fce12cfba0244b95752fd1bf4bb3;origin=https://github.com/amato-gianluca/UniMath;visit=swh:1:snp:c840c96e2c4f3571fd3082b82697cbaaac480f2c;anchor=swh:1:rel:1db4fad6810e17a5a9dcaa0563a43b566aaca252;path=/UniMath/Algebra/Universal/VTerms.v;lines=87-91}\swhl{\texttt{build\_term'}} is just the curried version of \lncode{build_term}.} that will be used to specify the equations of the theory of monoids.
\preparalink{https://archive.softwareheritage.org/swh:1:cnt:3a5262a9e829e569e458851f429af03dee73e634;origin=https://github.com/amato-gianluca/UniMath;visit=swh:1:snp:c840c96e2c4f3571fd3082b82697cbaaac480f2c;anchor=swh:1:rel:1db4fad6810e17a5a9dcaa0563a43b566aaca252;path=/UniMath/Algebra/Universal/Examples/Monoid.v;lines=33-34}
\begin{code}
  Definition |\swhl{monoid\_varspec}|: varspec monoid_signature
    := make_varspec monoid_signature natset (λ _, tt).
\end{code}
\preparalink{https://archive.softwareheritage.org/swh:1:cnt:3a5262a9e829e569e458851f429af03dee73e634;origin=https://github.com/amato-gianluca/UniMath;visit=swh:1:snp:c840c96e2c4f3571fd3082b82697cbaaac480f2c;anchor=swh:1:rel:1db4fad6810e17a5a9dcaa0563a43b566aaca252;path=/UniMath/Algebra/Universal/Examples/Monoid.v;lines=36}
\begin{code}
  Definition |\swhl{Mon}|: UU := term monoid_signature monoid_varspec tt.
\end{code}
\preparalink{https://archive.softwareheritage.org/swh:1:cnt:3a5262a9e829e569e458851f429af03dee73e634;origin=https://github.com/amato-gianluca/UniMath;visit=swh:1:snp:c840c96e2c4f3571fd3082b82697cbaaac480f2c;anchor=swh:1:rel:1db4fad6810e17a5a9dcaa0563a43b566aaca252;path=/UniMath/Algebra/Universal/Examples/Monoid.v;lines=37}
\begin{code}
  Definition |\swhl{mul}|: Mon → Mon → Mon := build_term' (●0: names monoid_signature).
\end{code}
\preparalink{https://archive.softwareheritage.org/swh:1:cnt:3a5262a9e829e569e458851f429af03dee73e634;origin=https://github.com/amato-gianluca/UniMath;visit=swh:1:snp:c840c96e2c4f3571fd3082b82697cbaaac480f2c;anchor=swh:1:rel:1db4fad6810e17a5a9dcaa0563a43b566aaca252;path=/UniMath/Algebra/Universal/Examples/Monoid.v;lines=38}
\begin{code}
  Definition |\swhl{id}|: Mon := build_term' (●1: names monoid_signature).
\end{code}
Term variables are associated to natural numebers.  In this case, three variables \preparalink{https://archive.softwareheritage.org/swh:1:cnt:3a5262a9e829e569e458851f429af03dee73e634;origin=https://github.com/amato-gianluca/UniMath;visit=swh:1:snp:c840c96e2c4f3571fd3082b82697cbaaac480f2c;anchor=swh:1:rel:1db4fad6810e17a5a9dcaa0563a43b566aaca252;path=/UniMath/Algebra/Universal/Examples/Monoid.v;lines=40-42}\lncode{|\swhl{x, y, z}|} will suffice for our needs:
\begin{code}
  Definition x : Mon := varterm (0: monoid_varspec).
  Definition y : Mon := varterm (1: monoid_varspec).
  Definition z : Mon := varterm (2: monoid_varspec).
\end{code}
Now, we have all the ingredients to specify our equations: the monoid axioms of associativity, left identity, and right identity (where \preparalink{https://archive.softwareheritage.org/swh:1:cnt:e759d6f54ecfb5252e1e288cfd605bb7a0fd35c9;origin=https://github.com/amato-gianluca/UniMath;visit=swh:1:snp:c840c96e2c4f3571fd3082b82697cbaaac480f2c;anchor=swh:1:rel:1db4fad6810e17a5a9dcaa0563a43b566aaca252;path=/UniMath/Algebra/Universal/Equations.v;lines=36}\lncode{|\swhl{==}|} is just a shorthand for giving terms of the relevant type \lncode{equation monoid_signature monoid_varspec}).
\preparalink{https://archive.softwareheritage.org/swh:1:cnt:3a5262a9e829e569e458851f429af03dee73e634;origin=https://github.com/amato-gianluca/UniMath;visit=swh:1:snp:c840c96e2c4f3571fd3082b82697cbaaac480f2c;anchor=swh:1:rel:1db4fad6810e17a5a9dcaa0563a43b566aaca252;path=/UniMath/Algebra/Universal/Examples/Monoid.v;lines=44}
\begin{code}
  Definition |\swhl{monoid\_mul\_assoc}| := mul (mul x y) z == mul x (mul y z).
\end{code}
\preparalink{https://archive.softwareheritage.org/swh:1:cnt:3a5262a9e829e569e458851f429af03dee73e634;origin=https://github.com/amato-gianluca/UniMath;visit=swh:1:snp:c840c96e2c4f3571fd3082b82697cbaaac480f2c;anchor=swh:1:rel:1db4fad6810e17a5a9dcaa0563a43b566aaca252;path=/UniMath/Algebra/Universal/Examples/Monoid.v;lines=45}
\begin{code}
  Definition |\swhl{monoid\_mul\_lid}| := mul id x == x.
\end{code}
\preparalink{https://archive.softwareheritage.org/swh:1:cnt:3a5262a9e829e569e458851f429af03dee73e634;origin=https://github.com/amato-gianluca/UniMath;visit=swh:1:snp:c840c96e2c4f3571fd3082b82697cbaaac480f2c;anchor=swh:1:rel:1db4fad6810e17a5a9dcaa0563a43b566aaca252;path=/UniMath/Algebra/Universal/Examples/Monoid.v;lines=46}
\begin{code}
  Definition |\swhl{monoid\_mul\_rid}| := mul x id == x.
\end{code}
We pack the above equations together into an equation system (\preparalink{https://archive.softwareheritage.org/swh:1:cnt:3a5262a9e829e569e458851f429af03dee73e634;origin=https://github.com/amato-gianluca/UniMath;visit=swh:1:snp:c840c96e2c4f3571fd3082b82697cbaaac480f2c;anchor=swh:1:rel:1db4fad6810e17a5a9dcaa0563a43b566aaca252;path=/UniMath/Algebra/Universal/Examples/Monoid.v;lines=48-49}\lncode{|\swhl{monoid\_axioms}|}) and its associated equational specification (\preparalink{https://archive.softwareheritage.org/swh:1:cnt:3a5262a9e829e569e458851f429af03dee73e634;origin=https://github.com/amato-gianluca/UniMath;visit=swh:1:snp:c840c96e2c4f3571fd3082b82697cbaaac480f2c;anchor=swh:1:rel:1db4fad6810e17a5a9dcaa0563a43b566aaca252;path=/UniMath/Algebra/Universal/Examples/Monoid.v;lines=51-56}\lncode{|\swhl{monoid\_eqspec}|}); finally, we define the class of equational algebras of monoids \preparalink{https://archive.softwareheritage.org/swh:1:cnt:3a5262a9e829e569e458851f429af03dee73e634;origin=https://github.com/amato-gianluca/UniMath;visit=swh:1:snp:c840c96e2c4f3571fd3082b82697cbaaac480f2c;anchor=swh:1:rel:1db4fad6810e17a5a9dcaa0563a43b566aaca252;path=/UniMath/Algebra/Universal/Examples/Monoid.v;lines=58}\lncode{|\swhl{monoid\_eqalgebra}|}.\footnote{We omit the formal construction which is uncomplicated and essentially reduces to uninteresting bookkeeping.}

Next, we want to show that every ``classical'' monoid $M$ has a natural structure of equational algebra.

We have two show that $M$ is a model for our equation system. Let us consider the left-identity axiom
\preparalink{https://archive.softwareheritage.org/swh:1:cnt:3a5262a9e829e569e458851f429af03dee73e634;origin=https://github.com/amato-gianluca/UniMath;visit=swh:1:snp:c840c96e2c4f3571fd3082b82697cbaaac480f2c;anchor=swh:1:rel:1db4fad6810e17a5a9dcaa0563a43b566aaca252;path=/UniMath/Algebra/Universal/Examples/Monoid.v;lines=66-72}
\begin{code}
  Lemma |\swhl{holds\_monoid\_mul\_lid}| : holds (monoid_algebra M) monoid_mul_lid.
  Proof.
    intro α. cbn in α.
    change (fromterm (monoid_algebra M) α tt (mul id x) = α 0).
    change (op (unel M) (α 0) = α 0).
    apply lunax.
  Qed.
\end{code}
As you see, we fix the variable evaluation \lncode{α}, then we observe that our goal reduces to the same law expressed in the usual language of monoids -- \lncode{op} for the product, \lncode{unel M} for the identity, \lncode{α 0} for the first variable \lncode{x} -- and then the goal is solved at once by applying the corresponding monoid axiom \preparalink{https://archive.softwareheritage.org/swh:1:cnt:51ab805f9977b358fc9de10d20558015c90ad566;origin=https://github.com/amato-gianluca/UniMath;visit=swh:1:snp:c840c96e2c4f3571fd3082b82697cbaaac480f2c;anchor=swh:1:rel:1db4fad6810e17a5a9dcaa0563a43b566aaca252;path=/UniMath/Algebra/Monoids.v;lines=68}\lncode{|\swhl{lunax}|}.

The other two laws -- for \preparalink{https://archive.softwareheritage.org/swh:1:cnt:3a5262a9e829e569e458851f429af03dee73e634;origin=https://github.com/amato-gianluca/UniMath;visit=swh:1:snp:c840c96e2c4f3571fd3082b82697cbaaac480f2c;anchor=swh:1:rel:1db4fad6810e17a5a9dcaa0563a43b566aaca252;path=/UniMath/Algebra/Universal/Examples/Monoid.v;lines=74-79}\swhl{right identity} and \preparalink{https://archive.softwareheritage.org/swh:1:cnt:3a5262a9e829e569e458851f429af03dee73e634;origin=https://github.com/amato-gianluca/UniMath;visit=swh:1:snp:c840c96e2c4f3571fd3082b82697cbaaac480f2c;anchor=swh:1:rel:1db4fad6810e17a5a9dcaa0563a43b566aaca252;path=/UniMath/Algebra/Universal/Examples/Monoid.v;lines=81-86}\swhl{associativity} -- are proven in the same way.

Thus, we can now pack everything into a monoid eqalgebra with \preparalink{https://archive.softwareheritage.org/swh:1:cnt:3a5262a9e829e569e458851f429af03dee73e634;origin=https://github.com/amato-gianluca/UniMath;visit=swh:1:snp:c840c96e2c4f3571fd3082b82697cbaaac480f2c;anchor=swh:1:rel:1db4fad6810e17a5a9dcaa0563a43b566aaca252;path=/UniMath/Algebra/Universal/Examples/Monoid.v;lines=88-96}\lncode{|\swhl{is\_eqalgebra\_monoid}|} and \preparalink{https://archive.softwareheritage.org/swh:1:cnt:3a5262a9e829e569e458851f429af03dee73e634;origin=https://github.com/amato-gianluca/UniMath;visit=swh:1:snp:c840c96e2c4f3571fd3082b82697cbaaac480f2c;anchor=swh:1:rel:1db4fad6810e17a5a9dcaa0563a43b566aaca252;path=/UniMath/Algebra/Universal/Examples/Monoid.v;lines=98-102}\lncode{|\swhl{make\_monoid\_eqalgebra}|}.

\subsection{Algebra of booleans and Tarski's semantics}\label{sec:tarski}

We conclude the code survey with a further example based on a simple single sorted algebraic language: the algebra of booleans, and its connectives.\footnote{The code for this example can be found in the module \preparalink{https://archive.softwareheritage.org/swh:1:cnt:3a51e836749a6f3da9b147bdb4af6eb0575a3507;origin=https://github.com/amato-gianluca/UniMath;visit=swh:1:snp:c840c96e2c4f3571fd3082b82697cbaaac480f2c;anchor=swh:1:rel:1db4fad6810e17a5a9dcaa0563a43b566aaca252;path=/UniMath/Algebra/Universal/Examples/Bool.v}\swhl{\texttt{Bool.v}}
}

\vspace{.28cm}

The language considered has the usual boolean operators: truth, falsity, negation, conjunction, disjunction, and implication.
Arities can be simply specified by naturals (the number of arguments).

We use the function \lncode{make_signature_simple_single_sorted} to build a signature from the list of arities:
\preparalink{https://archive.softwareheritage.org/swh:1:cnt:3a51e836749a6f3da9b147bdb4af6eb0575a3507;origin=https://github.com/amato-gianluca/UniMath;visit=swh:1:snp:c840c96e2c4f3571fd3082b82697cbaaac480f2c;anchor=swh:1:rel:1db4fad6810e17a5a9dcaa0563a43b566aaca252;path=/UniMath/Algebra/Universal/Examples/Bool.v;lines=17}
\begin{code}
  Definition |\swhl{bool\_signature}| :=
    make_signature_simple_single_sorted [0; 0; 1; 2; 2; 2].
\end{code}
Obviously, the type of booleans is already defined in UniMath, together with its usual constants and operations: \preparalink{https://archive.softwareheritage.org/swh:1:cnt:fd4f9df69c54bf298ff64a07922a36fc030b1dc4;origin=https://github.com/amato-gianluca/UniMath;visit=swh:1:snp:c840c96e2c4f3571fd3082b82697cbaaac480f2c;anchor=swh:1:rel:1db4fad6810e17a5a9dcaa0563a43b566aaca252;path=/UniMath/Foundations/Preamble.v;lines=35}\lncode{|\swhl{false}|}, \preparalink{https://archive.softwareheritage.org/swh:1:cnt:fd4f9df69c54bf298ff64a07922a36fc030b1dc4;origin=https://github.com/amato-gianluca/UniMath;visit=swh:1:snp:c840c96e2c4f3571fd3082b82697cbaaac480f2c;anchor=swh:1:rel:1db4fad6810e17a5a9dcaa0563a43b566aaca252;path=/UniMath/Foundations/Preamble.v;lines=34}\lncode{|\swhl{true}|}, \preparalink{https://archive.softwareheritage.org/swh:1:cnt:fd4f9df69c54bf298ff64a07922a36fc030b1dc4;origin=https://github.com/amato-gianluca/UniMath;visit=swh:1:snp:c840c96e2c4f3571fd3082b82697cbaaac480f2c;anchor=swh:1:rel:1db4fad6810e17a5a9dcaa0563a43b566aaca252;path=/UniMath/Foundations/Preamble.v;lines=37}\lncode{|\swhl{negb}|}, \preparalink{https://archive.softwareheritage.org/swh:1:cnt:ea8de322e62bb29fc8a4250d4a7920d216448e1e;origin=https://github.com/amato-gianluca/UniMath;visit=swh:1:snp:c840c96e2c4f3571fd3082b82697cbaaac480f2c;anchor=swh:1:rel:1db4fad6810e17a5a9dcaa0563a43b566aaca252;path=/UniMath/MoreFoundations/Bool.v;lines=6-9}\lncode{|\swhl{andb}|}, \preparalink{https://archive.softwareheritage.org/swh:1:cnt:ea8de322e62bb29fc8a4250d4a7920d216448e1e;origin=https://github.com/amato-gianluca/UniMath;visit=swh:1:snp:c840c96e2c4f3571fd3082b82697cbaaac480f2c;anchor=swh:1:rel:1db4fad6810e17a5a9dcaa0563a43b566aaca252;path=/UniMath/MoreFoundations/Bool.v;lines=11-14}\lncode{|\swhl{orb}|}, \preparalink{https://archive.softwareheritage.org/swh:1:cnt:ea8de322e62bb29fc8a4250d4a7920d216448e1e;origin=https://github.com/amato-gianluca/UniMath;visit=swh:1:snp:c840c96e2c4f3571fd3082b82697cbaaac480f2c;anchor=swh:1:rel:1db4fad6810e17a5a9dcaa0563a43b566aaca252;path=/UniMath/MoreFoundations/Bool.v;lines=16-19}\lncode{|\swhl{implb}|}.

Now, booleans form an hSet, which is denoted \preparalink{https://archive.softwareheritage.org/swh:1:cnt:bdeb7d0a286b08529ec4d7a5c15edc6c3f855141;origin=https://github.com/amato-gianluca/UniMath;visit=swh:1:snp:c840c96e2c4f3571fd3082b82697cbaaac480f2c;anchor=swh:1:rel:1db4fad6810e17a5a9dcaa0563a43b566aaca252;path=/UniMath/Foundations/Sets.v;lines=163}\lncode{|\swhl{boolset}|}.
It is easy to organize all of those constituents into an algebra for our signature by specifying the translation:
\preparalink{https://archive.softwareheritage.org/swh:1:cnt:3a51e836749a6f3da9b147bdb4af6eb0575a3507;origin=https://github.com/amato-gianluca/UniMath;visit=swh:1:snp:c840c96e2c4f3571fd3082b82697cbaaac480f2c;anchor=swh:1:rel:1db4fad6810e17a5a9dcaa0563a43b566aaca252;path=/UniMath/Algebra/Universal/Examples/Bool.v;lines=21-24}
\begin{code}
  Definition |\swhl{bool\_algebra}| := make_algebra_simple_single_sorted' 
    bool_signature boolset
    [( false ; true ; negb ; andb ; orb ; implb )].
\end{code}
Next, we build the algebra of (open) terms, that is, boolean formulas.

This is done in two steps. First, we give a variable specification, i.e.~a set of type variables:
\preparalink{https://archive.softwareheritage.org/swh:1:cnt:3a51e836749a6f3da9b147bdb4af6eb0575a3507;origin=https://github.com/amato-gianluca/UniMath;visit=swh:1:snp:c840c96e2c4f3571fd3082b82697cbaaac480f2c;anchor=swh:1:rel:1db4fad6810e17a5a9dcaa0563a43b566aaca252;path=/UniMath/Algebra/Universal/Examples/Bool.v;lines=51}
\begin{code}
  Definition |\swhl{bool\_varspec}| := make_varspec bool_signature natset (λ _, tt).
\end{code}
Then, we define the algebra of terms and the associated constructors.
\preparalink{https://archive.softwareheritage.org/swh:1:cnt:3a51e836749a6f3da9b147bdb4af6eb0575a3507;origin=https://github.com/amato-gianluca/UniMath;visit=swh:1:snp:c840c96e2c4f3571fd3082b82697cbaaac480f2c;anchor=swh:1:rel:1db4fad6810e17a5a9dcaa0563a43b566aaca252;path=/UniMath/Algebra/Universal/Examples/Bool.v;lines=55}
\begin{code}
  Definition T := term bool_signature bool_varspec tt.
\end{code}
\preparalink{https://archive.softwareheritage.org/swh:1:cnt:3a51e836749a6f3da9b147bdb4af6eb0575a3507;origin=https://github.com/amato-gianluca/UniMath;visit=swh:1:snp:c840c96e2c4f3571fd3082b82697cbaaac480f2c;anchor=swh:1:rel:1db4fad6810e17a5a9dcaa0563a43b566aaca252;path=/UniMath/Algebra/Universal/Examples/Bool.v;lines=59}
\begin{code}
  Definition |\swhl{bot}|  : T         := build_term' (●0 : names bool_signature).
\end{code}
\preparalink{https://archive.softwareheritage.org/swh:1:cnt:3a51e836749a6f3da9b147bdb4af6eb0575a3507;origin=https://github.com/amato-gianluca/UniMath;visit=swh:1:snp:c840c96e2c4f3571fd3082b82697cbaaac480f2c;anchor=swh:1:rel:1db4fad6810e17a5a9dcaa0563a43b566aaca252;path=/UniMath/Algebra/Universal/Examples/Bool.v;lines=60}
\begin{code}
  Definition |\swhl{top}|  : T         := build_term' (●1 : names bool_signature).
\end{code}
\preparalink{https://archive.softwareheritage.org/swh:1:cnt:3a51e836749a6f3da9b147bdb4af6eb0575a3507;origin=https://github.com/amato-gianluca/UniMath;visit=swh:1:snp:c840c96e2c4f3571fd3082b82697cbaaac480f2c;anchor=swh:1:rel:1db4fad6810e17a5a9dcaa0563a43b566aaca252;path=/UniMath/Algebra/Universal/Examples/Bool.v;lines=61}
\begin{code}
  Definition |\swhl{neg}|  : T → T     := build_term' (●2 : names bool_signature).
\end{code}
\preparalink{https://archive.softwareheritage.org/swh:1:cnt:3a51e836749a6f3da9b147bdb4af6eb0575a3507;origin=https://github.com/amato-gianluca/UniMath;visit=swh:1:snp:c840c96e2c4f3571fd3082b82697cbaaac480f2c;anchor=swh:1:rel:1db4fad6810e17a5a9dcaa0563a43b566aaca252;path=/UniMath/Algebra/Universal/Examples/Bool.v;lines=62}
\begin{code}
  Definition |\swhl{conj}| : T → T → T := build_term' (●3 : names bool_signature).
\end{code}
\preparalink{https://archive.softwareheritage.org/swh:1:cnt:3a51e836749a6f3da9b147bdb4af6eb0575a3507;origin=https://github.com/amato-gianluca/UniMath;visit=swh:1:snp:c840c96e2c4f3571fd3082b82697cbaaac480f2c;anchor=swh:1:rel:1db4fad6810e17a5a9dcaa0563a43b566aaca252;path=/UniMath/Algebra/Universal/Examples/Bool.v;lines=63}
\begin{code}
  Definition |\swhl{disj}| : T → T → T := build_term' (●4 : names bool_signature).
\end{code}
\preparalink{https://archive.softwareheritage.org/swh:1:cnt:3a51e836749a6f3da9b147bdb4af6eb0575a3507;origin=https://github.com/amato-gianluca/UniMath;visit=swh:1:snp:c840c96e2c4f3571fd3082b82697cbaaac480f2c;anchor=swh:1:rel:1db4fad6810e17a5a9dcaa0563a43b566aaca252;path=/UniMath/Algebra/Universal/Examples/Bool.v;lines=64}
\begin{code}
  Definition |\swhl{impl}| : T → T → T := build_term' (●5 : names bool_signature).
\end{code}
Finally, we use the universal property of the term algebra to define the interpretation of boolean formulas:
\preparalink{https://archive.softwareheritage.org/swh:1:cnt:3a51e836749a6f3da9b147bdb4af6eb0575a3507;origin=https://github.com/amato-gianluca/UniMath;visit=swh:1:snp:c840c96e2c4f3571fd3082b82697cbaaac480f2c;anchor=swh:1:rel:1db4fad6810e17a5a9dcaa0563a43b566aaca252;path=/UniMath/Algebra/Universal/Examples/Bool.v;lines=68-69}
\begin{code}
  Definition |\swhl{interp}| (α: assignment bool_algebra bool_varspec) (t: T) : bool :=
    fromterm (ops bool_algebra) α tt t.
\end{code}
At this point, we can check the effectiveness of our definitions with some applications.

To set-up our tests, we introduce three variables \preparalink{https://archive.softwareheritage.org/swh:1:cnt:3a51e836749a6f3da9b147bdb4af6eb0575a3507;origin=https://github.com/amato-gianluca/UniMath;visit=swh:1:snp:c840c96e2c4f3571fd3082b82697cbaaac480f2c;anchor=swh:1:rel:1db4fad6810e17a5a9dcaa0563a43b566aaca252;path=/UniMath/Algebra/Universal/Examples/Bool.v;lines=75-77}\lncode{|\swhl{x, y, z}|}
and a simple evaluation function \preparalink{https://archive.softwareheritage.org/swh:1:cnt:3a51e836749a6f3da9b147bdb4af6eb0575a3507;origin=https://github.com/amato-gianluca/UniMath;visit=swh:1:snp:c840c96e2c4f3571fd3082b82697cbaaac480f2c;anchor=swh:1:rel:1db4fad6810e17a5a9dcaa0563a43b566aaca252;path=/UniMath/Algebra/Universal/Examples/Bool.v;lines=88-93}\lncode{|\swhl{v}|} for variables that assigns \lncode{true} to the variable \lncode{x} and \lncode{y} (the variable of index $0$ and $1$) and \lncode{false} otherwise.
Now, we can interpret formulas such as $x \wedge (z \to \neg y)$:
\begin{code}
  Goal interp v (conj x (conj (neg  y) z)) = false.
  Proof. lazy. apply idpath. Qed.
\end{code}
The reader is invited to notice that the choice of the lazy strategy is not accidental. Computations required to evaluate such a proof term are pretty heavy and the standard call by value strategy does not seem able to produce a result in reasonable time.

\vspace{.28cm}

A few other examples are available in our code as, for instance, a proof of Dummett's tautology:
\preparalink{https://archive.softwareheritage.org/swh:1:cnt:3a51e836749a6f3da9b147bdb4af6eb0575a3507;origin=https://github.com/amato-gianluca/UniMath;visit=swh:1:snp:c840c96e2c4f3571fd3082b82697cbaaac480f2c;anchor=swh:1:rel:1db4fad6810e17a5a9dcaa0563a43b566aaca252;path=/UniMath/Algebra/Universal/Examples/Bool.v;lines=105-111}
\begin{code}
  Lemma |\swhl{Dummett}| : ∏ i, interp i (disj (impl x y) (impl y x)) = true.
  Proof.
    intro i. lazy.
    induction (i 0); induction (i 1); apply idpath.
  Qed.
\end{code}

Notice that this formal proof is just a case analysis for truth-tables in disguise: we instantiate the values of \lncode{x} and \lncode{y} by applying \lncode{induction} twice, but the remaining job is left to the computing mechanism of Coq, which is able to autonomously verify that the evaluation does yield the value \lncode{true} in all cases -- we only need to apply \lncode{idpath}.

\section*{Conclusions}\label{works}
\addcontentsline{toc}{section}{Conclusions}

We have surveyed our UniMath library for universal algebra, covering the fundamental concepts of multi-sorted signatures, algebras, and equational algebras.

We have shown how to implement term algebras over a signature without relying on general inductive constructions (as prescribed by the UniMath formal language), and proven that our single sorted ground term algebras have the structure of homotopy W-types.

Additionally, we showed that algebras (and algebras modulo equations) over a given signature define a univalent category, whenever their carriers are (sorted) hSets.

Finally, we have instantiated with three concrete examples of algebraic structures our general framework for formalising universal algebra in UniMath.

\subsection*{Future work}
We plan to enhance our implementation along three directions:
\begin{enumerate}
\item First of all, we wish to streamline the interface provided by the library. With the current state of implementation, the user is exposed to many technical details which have no theoretical relevance. These include the internal signatures generated by \lncode{vsignature} for dealing with variables in terms and the existence of two term algebras, one for ground terms, the other for general terms, while the former should only be a particular case of the latter. 

We plan to redesign the interface in order to hide the internal details as much as possible. Furthermore, the interface for heterogeneous vectors might be generalized to make the \lncode{HVectors.v} module more useful outside of the scope of our library.
\item
Next, we intend to push further the mathematical salience of the library. This means, for instance, generalising the result relating ground term algebras and W-types to the multi sorted case. Since we expect to require \emph{indexed} homotopy W-types to keep track of the sorts, we aim at providing UniMath with a detailed formalisation of that notion, following the analysis given by~\cite{Sattler2015OnRI}. Moreover, we plan to complete the treatment of equational algebras by defining the initial algebra of terms modulo equational congruence. Additionally, we expect to give formal proofs of some known central results in universal algebra, starting from the homomorphism theorems and Birkhoff's variety theorem.
\item
Finally, to extend the library with refined applications and examples of univalent reasoning. This would give evidence that even the minimalist environment of UniMath does allow its user to approach mechanised mathematics with the advantages of both univalent reasoning -- to handle equivalent objects as naturally as in informal mathematics -- and the automation process of the proof assistant -- to be smartly used for performing ``internal'' implementations in order to leave all computations with no demonstrative significance to the machine.
\end{enumerate}

\subsection*{Related work}
Ours is not the first mechanisation of universal algebra currently available in the literature.

A classical work on formalising this area of mathematics in dependent type theory is \cite{capretta:1999}, where he systematically uses setoids in Coq to handle equality on structures.  Another attempt, still based on setoids, has been carried out in Agda by \cite{gunther-gadea-pagano:2018}.

The works of \cite{demeo2021agda, demeo2021agda2, demeo2021machine} draw on the multi-sorted version of \cite{DBLP:journals/corr/abs-2111-07936} to develop an extensive and setoid-based Agda library on single-sorted universal algebra that strives to be as powerful as Abel's formalisation but a bit more sensitive to foundational aspects.

Very recently, the paper~\cite{REYNOLDS2024103054} presents a formalisation in Coq of the theory of institutions based on the approach to universal algebra by~\cite{gunther-gadea-pagano:2018}.

\medskip

On the categorical side, initial semantics furnishes elegant techniques for studying induction and recursion principles within a general algebraic setting with applications in programming languages and logic. Assuming univalence, steady research activity has produced over the time a number of contributions to the UniMath library, see e.g.~\cite{ahrens_et_al:LIPIcs:2018:9671, DBLP:journals/jar/AhrensMM19, 10.1145/3497775.3503678}.

From a HOTT-UF perspective, the formalisation of universal algebra by \cite{lynge:2017} -- further developed in \cite{lynge-spitters:2019} --  outline a framework that more closely compares with ours, since it is still based on univalent reasoning, though implemented in the Coq-HoTT~\citep{DBLP:conf/cpp/BauerGLSSS17} extension for the Calculus of the (Co)Inductive Constructions.

\section*{Acknowledgements}
The authors thank the \emph{European Research Network on Formal Proofs} (COST Action EuroProofNet CA20111) and the \emph{European research network on types for programming and verification} (COST Action EUTypes CA15123) for supporting the School and Workshop on Univalent Mathematics 2019 where this project had its origin.

GA's work is partially supported by the Gruppo Nazionale per il Calcolo Scientifico (GNCS) of the Istituto Nazionale di Alta Matematica (INdAM).

MC and MM's work is partially supported by the Italian Ministry of University and Research and the Gruppo Nazionale per le Strutture Algebriche, Geometriche e le loro Applicazioni (GNSAGA) of the Istituto Nazionale di Alta Matematica (INdAM).

CPB's work is partially supported by the project SERICS (PE00000014) under the MUR National Recovery and Resilience Plan funded by the European Union -- NextGenerationEU.

\bibliography{biblio2}

\end{document}